\documentclass[twocolumn,showkeys,pra]{revtex4}

\usepackage[utf8x]{inputenc}
\usepackage{graphicx}
\usepackage{epstopdf}
\usepackage{epsfig}
\usepackage{amssymb}
\usepackage{setspace}
\usepackage{amsmath}
\usepackage{textcomp}
\usepackage{url}
\usepackage[pdftex]{color}
\usepackage{subfigure}
\usepackage{physics}
\usepackage{dsfont}
\usepackage{dcolumn}
\usepackage[T1]{fontenc}
\usepackage{mathptmx}
\usepackage{etoolbox}
\usepackage{cancel}
\usepackage{ulem}

\begin{document}

\title{Unraveling Rodeo Algorithm Through the Zeeman Model} 

\author{R. F. I. Gomes}
\affiliation{Universidade Federal da Integração Latino-Americana, Foz do Iguaçu, Paraná, Brasil.}

\author{J. S. C. Rocha}
\affiliation{Departamento de Física, Universidade Federal de Ouro Preto - UFOP, Ouro Preto, Minas Gerais, Brasil.}

\author{W. A. T. Nogueira}
\affiliation{Departamento de Física, Universidade Federal de Juiz de Fora - UFJF, Minas Gerais, Brasil.}

\author{R. A. Dias}
\affiliation{Departamento de Física, Universidade Federal de Juiz de Fora - UFJF, Minas Gerais, Brasil.}
\begin{abstract}

We unravel the Rodeo Algorithm to determine the eigenstates and eigenvalues spectrum for a general Hamiltonian considering arbitrary initial states. By presenting a novel methodology, we detail the original method and show how to define all properties without having prior knowledge regarding the eigenstates. To this end, we exploit Pennylane and Qiskit platforms resources to analyze scenarios where the Hamiltonians are described by the Zeeman model for one and two spins. We also introduce strategies and techniques to improve the algorithm's performance by adjusting its intrinsic parameters and reducing the fluctuations inherent to data distribution. First, we explore the dynamics of a single qubit on Xanadu simulators to set the parameters that optimize the method performance and select the best strategies to execute the algorithm. On the sequence, we extend the methodology for bipartite systems to discuss how the algorithm works
when degeneracy and entanglement are taken into account. Finally, we compare the predictions with the results obtained on a real superconducting device provided by the IBM Q Experience program, establishing the conditions to increase the protocol efficiency for multi-qubit systems.
\end{abstract}


\maketitle

\section{Introduction}
\label{sec:introduction}

One of the primary challenges in quantum mechanics consists of characterizing a quantum system and the corresponding measurable values of its observables, such as energy or momentum \cite{cohen2019book}. In this sense, it is crucial to adopt a suitable procedure to predict  the system's temporal evolution and reduce the time expended to reach this information   
\cite{sakurai2017book,nielsen2010book,rigol2008}, since solving the eigenstate problem - especially for intricate systems with numerous particles or degrees of freedom - is often presented as an arduous task. 
\cite{dirac1981book,modisette1996,polizzi2009,beugeling2018,schutt2019}. In cases where an analytical solution or even an experimental implementation of the system is not feasible, measurements over a virtual model can be a reliable alternative \cite{landau2021book}. In this regard, quantum computation offers a natural tool to address this problem, as it is based on strictly quantum phenomena (such as superposition and entanglement) with no counterpart in classical computation \cite{feynman1982,lloyd1996,lloyd1998,abrams1999,buluta2009,brown2010,aaronson2011,georgescu2014,cerezo2021}.

Fortunately, it is possible to build and simulate specific codes with quantum states through Noisy Intermediate-Scale Quantum (NISQ) devices, whose name was proposed by John Preskill \cite{preskill2018} to describe the era of “short-term” quantum computing. Although there is still no quantum computer capable of operating with a large amount of fully connected qubits without being susceptible to decoherence (and with a comprehensive error correction protocol), these devices can be used to compare whether there is any quantum advantage over classic methods and execute algorithms combined with proper techniques such as VQE \cite{peruzzo2014} and QAOA \cite{farhi2014}. 

In this context, Choi et al. have recently introduced a method called Rodeo Algorithm \cite{choi2021} to find the solutions for the time-independent Schr\"odinger equation ($H\ket{\psi} = E\ket{\psi}$)
by solving the eigenstates and eigenvalues problem of an arbitrary Hamiltonian operator.\,\,Compared to previous approaches such as phase estimation and adiabatic evolution, it is expected that this algorithm provides the desired results in an exponentially faster timescale \cite{choi2021}. Based on it, some applications have been proposed in {\cite{sherbert2022,guzman2022,pederiva2021, perez2022, stetcu2022, qian2024, cohen2023, lindgren2024}.} 

Rodeo Algorithm exploits the phase kickback phenomenon \cite{nielsen2010book}, which can be considered as a shift from the unitary phase associated to the qubit target dynamics to the ancilla that controls its evolution. The method proceeds as follows: first, a set of $N$ ancillary qubits are prepared in the state $\ket{1}^{\otimes N}$ and an initial pure state $\ket{\psi_I}$ is inserted into the circuit. On the sequence, each of the $k^{th}$ ancilla is rotated to $\ket{-}=(\ket{0}-\ket{1})/{\sqrt{2}}$ and connected to $\ket{\psi_I}$ to control the time evolution operator
$O = \exp(-iH_{\text{obj}}t_k)$, where $H_{\text{obj}}$ is an arbitrary Hamiltonian that dictates the quantum system dynamics for a random time interval $t_k$. Next, a variable $E$ (also random) is assigned as a trial to one of the eigenvalues related to $H_{\text{obj}}$. The choice is followed by a phase shift operator $P = \exp(iEt_k)$ applied to each ancilla, which are submitted to another Hadamard gate and measured on the computational basis (defined as the $\sigma_{z}$ operator eigenstates).  

\begin{figure}[!ht]
    \includegraphics[scale=0.4]{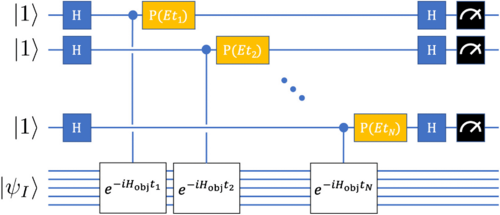}
    \caption{Circuit diagram for Rodeo Algorithm. The target system is represented by $|\psi_I⟩$, while all ancilla are initialized in the state $|1⟩$ and rotate to $|-⟩$ by a Hadamard gate ($H$). This set is used to control the time evolution of the object Hamiltonian ($H_{\text{obj}}$) for random values of $t_k$ associated with each qubit labeled by $k=1,…,N$. The operation is followed by a phase rotation ($P(E t_k)$), another Hadamard gate ($H$), and a final measurement on the computational basis. Adapted from \cite{choi2021}.}
    \label{fig:RodeoPRL}
\end{figure}

The circuit diagram for the Rodeo Algorithm is shown in Fig.~\ref{fig:RodeoPRL}.
As discussed above, we can describe the action of $O$ over $\ket{\psi_I}$ by attaching the system's phase to the respective ancilla. Since the time evolution and the phase shift operators rotate the last in opposite directions, they will all collapse to $\ket{1}$ when $H_{\text{obj}}\ket{\psi_I} = E_{\text{obj}}\ket{\psi_I}$ and $E = E_{\text{obj}}$. Therefore, one can infer with high probability that the algorithm succeeds when the output registered for a measurement on the computational basis for each ancilla $\ket{k}$ is $\bra{k}\sigma_{z}\ket{k}\approx-1$, which leads to
\begin{eqnarray}
P_{N}=\prod_{k=1}^{N}
\cos^{2}{\left[(E_{\text{obj}}-E)\frac{t_{k}}{2}\right]} \approx 1.
\label{eq:probtotal}
\end{eqnarray}

In turn, there are some crucial limitations regarding the original purpose, since the authors state that adjusting the system's energy and controlling its temporal evolution is not enough to determine the spectrum of eigenstates and eigenvalues of $H_{\text{obj}}$. Furthermore, they show how to reach the eigenvalues by measuring the ancilla qubits, but it is not detailed the procedure to construct the set of eigenstates after that measurement. Moreover, it is argued that the success of the method depends largely on the overlap probability between the initial state and the eigenstates \cite{choi2021}, which implies that one must have prior information about the eigenstates and emphasizes the need to restrict the input to achieve both goals.

To overcome the restrictions concerning the dependence on previous information about the eigenstates - or at least part of them - and how to define their set, we apply the Rodeo Algorithm to a single and two-qubit Hamiltonians described by the Zemman model to determine its eigenvalues and eigenvectors spectrum with an arbitrary initial state. For this purpose, the capabilities of the Pennylane \cite{refpennylane} and Qiskit \cite{refqiskit} platforms are explored to analyze these scenarios. While Xanadu's systems allow to simulate circuits that scale up to 29 qubits (up to this date), the IBM Experience platform - despite being more limited in the number of qubits in open access accounts - provides the possibility to run the same codes on real devices. For all cases, we design strategies that go beyond these limits to optimize the algorithm and define the desired properties for any input.

This article is organized as follows: in section \ref{sec:bullOper} we introduce a new methodology through the ``Bull operator", which represents the sequence of transformations that happen onto the so-called ``Rider state" to implement Rodeo Algorithm. The general structure of the Zeeman model is presented in section \ref{sec:zeeman}, while section \ref{sec:dataset} explains how data obtained in the codes that support the results are organized. 
We discuss the strategies and techniques developed for the one and two-spin Zeeman model through the Pennylane simulator in section \ref{sec:results-sim}, whose results are compared with the ones obtained on an IBM superconducting device in section \ref{sec:results-dev}. Finally, we address our conclusions in section \ref{sec:conclusion}.
\section{The Bull Operator and The Rider State}
\label{sec:bullOper}

Let us now establish the Bull operator and the Rider state. The former will be defined as the controlled time evolution followed by the phase shift operator, whereas the latter represents the global state composed of the ancillary and target qubits. To generalize this approach, we consider that the circuit starts with a rank-$M$ state $\ket{\psi_I}$ and $N$ ancillary qubits. 

As mentioned in section \ref{sec:introduction}, each ancilla is initially set to $\ket{-}$, and we consider that the $M$ qubits required to store the target are randomly oriented to form a general state. Therefore, the Rider state can be initially written as
\begin{equation}
     \ket{\Psi_0} =  \frac{1}{\sqrt{2^N}}   \left[\bigotimes_{n=0}^{N-1}\left(\sum_{y_n=0}^1 e^{i\pi y_n} \ket{y_n}\right) \right] \otimes \ket{\psi_I},
     \label{eq:psi0}
\end{equation}
where $\ket{y_n}$ represents the standard basis ($\ket{0};\;\ket{1}$). Hence, all possible states spam the computational basis through the tensor products between the $N$ ancilla.

In turn, the Bull operator can be constructed as follows: first, note that the time evolution controlled by the $n^{th}$ ancilla is given by
\begin{align} 
\mathcal{CO}_n =  \sum_{x}\biggr[\mathds{1}^{\otimes (n-1)} & \otimes \left( \sum_{y_n=0}^{1} e^{-i E_x y_n t_n} \op{y_n}\right) \nonumber\\  
& \otimes \mathds{1}^{\otimes(N-n)} \biggr] \otimes  \op{x},
\label{eq:evolution}
\end{align}
where $\mathds{1}$ is the identity operator. Here, $\ket{x}$ represents a general vector from any eigenstate basis of the Halmitonian that acts over the target system.

On the other hand, the phase shift operation on the $n^{th}$ ancilla is described by the projector
\begin{equation}
\mathcal{P}_n =  \mathds{1}^{\otimes (n-1)}\otimes  \left( \sum_{y_n=0}^1 e^{i  E t_n y_n} \op{y_n}{y_n} \right) \otimes \mathds{1}^{\otimes (N-n + M)}.
\label{eq:phaseshift}
\end{equation}
From Eqs. (\ref{eq:evolution})  
and (\ref{eq:phaseshift}), 
we define $\mathcal{B}_n $ to implement the action of the cycle that twists these states side to side as
\begin{align} \nonumber
\mathcal{B}_n \equiv   \mathcal{P}_n\ \mathcal{CO}_n  =
   \sum_{x} \biggr[ \mathds{1}^{\otimes (n-1)} & \otimes \left( \sum_{y_n=0}^1 e^{i (E - E_x) t_n y_n} \op{y_n}\right) \\ & \otimes \mathds{1}^{\otimes (N-n)}\biggr] \otimes \op{x}.
\label{eq:evolutionphaseshift}
\end{align}

By considering that one ride occurs when $\mathcal{B}_n$ is performed sequentially over the $N$ ancillary qubits, the Bull operator results in
\begin{align} 
\mathcal{B}  &\equiv \prod_{n=0}^{N-1}\mathcal{B}_n \\
&= \sum_{x} \left[ \bigotimes_{n=0}^{N-1} \left( \sum_{y_n=0}^1 e^{i (E - E_x) t_n y_n} \op{y_n}\right)\right]\otimes \op{x},\nonumber 
\end{align}
taking into account that the eigenstates $\ket{x}$ forms an orthonormal basis and satisfies the condition $\bra{x'}\ket{x} = \delta(x,x')$. 
Consequently, the Rodeo Algorithm can be seen as the action of the Bull operator over the Rider state:
\begin{align}
 &\ket{\Psi_i} = \mathcal{B} \ket{\Psi_0} = &\\\nonumber &\frac{1}{\sqrt{2^N}} \sum_{x} \bra{x}\ket{\psi_I}  \left[  \bigotimes_{n=0}^{N-1}  \left( \sum_{y_n=0}^1 e^{i (E -E_x)(t_n + \pi)y_n} \ket{y_n}\right)\right] \otimes \ket{x}.&
 \end{align}

The subsequent step involves applying another Hadamard gate on each ancillary qubit. Thereby, the final global state is given by
\begin{align} 
&\ket{\Psi_f}=H^{\otimes N} \otimes   {\mathrm{1}^{\otimes M}}  \ket{\Psi_i}=&\nonumber\\
&\sum_{x}\bra{x} \ket{\psi_I} \left[  \bigotimes_{n=0}^{N-1} \left(  \frac{ 1 -  e^{i (E -E_x)t_n}}{2} \ket{0}   \right. \right. \nonumber\\
& \hspace{80pt} + \frac{ 1 + e^{i (E - E_x)t_n} }{2} \ket{1} \biggr)\Biggr]  \otimes  \ket{x}.
\label{eq:psi_f}
\end{align}

It must be pointed out that the global state $\ket{\Psi_f}$ is entangled, where the respective subsystems are composed of the ancillary and target qubits. Also, note that the description presented in Eq. (\ref{eq:psi_f}) is valid for any circuit configuration. 

The last step consists of measuring the expected value for each ancilla and comparing the result with the condition established in Eq. (\ref{eq:probtotal}). However, we decide to go further and calculate the average for the whole set according to \cite{rocha2023}:
\begin{eqnarray}
h(E,\psi_I|t) & = &\frac{1}{N} \sum_{j=1}^{N} \bra{\Psi_f} \mathds{1}^{\otimes j-1}\otimes\sigma_z^{j}\otimes \mathds{1}^{\otimes N-j+M} \ket{\Psi_f}.\nonumber\\
\label{eq:h_psi_f}
\end{eqnarray}
By replacing $\ket{\Psi_f}$ for the expression defined in Eq. (\ref{eq:psi_f}), this average results in 
\begin{equation}
h(E,\psi_I|t)  =  -\frac{1}{N}  \sum_{n=0}^{N-1} \sum_{x} |\bra{x}\ket{\psi_I}|^{2}   \cos{\left[ (E - E_x)t_n  \right]}.
\label{eq:h_final}
\end{equation}

At this point, we can use the result provided by Eq. (\ref{eq:h_final}) and calculate another average to extract more information about the system. Considering that the function $\overline{h}(E,\psi_I)$ will depend on the randomly sampled values $t_n$ generated by a given probability density function (PDF) $P(t|\tau,d)$, the result can be expressed as an integral over all possible values of $t$ weighted by the latter, i.e.
\begin{equation}
    \overline{h}(E,\psi_I) = \int^{\infty}_{-\infty} h(E,\psi_I|t) P(t|\tau,d) \mathrm{d}t.
\label{eq:MedGauss}
\end{equation}
Since the integration variable is generated by a PDF, Eq. (\ref{eq:MedGauss}) can be evaluated by a simple arithmetic mean in every importance sampling.

To conclude the procedure, we will follow the original method by setting the PDF as the normal distribution as
\begin{equation}
    P(t|\tau,d) = \frac{1}{d\sqrt{2\pi}} e^{-\frac{(t-\tau)^2}{2d^2}}.
\label{eq:GaussDist}
\end{equation}
By considering that
\begin{eqnarray}
I &=& 2 \frac{1}{d\sqrt{2\pi}}\int^{\infty}_{0} \cos(\alpha t) e^{-\frac{(t-\tau)^2}{2d^2}} \mathrm{d}t,\nonumber\\
  &=& e^{-d^2\alpha^2/2}\cos(\alpha\tau),
\end{eqnarray}
Eq. (\ref{eq:MedGauss}) results in
 \begin{eqnarray}  
     & &\overline{h}(E,\psi_I) \equiv \nonumber\\ & & -\sum_{x} |\bra{x}\ket{\psi_I}|^{2}   e^{-d^2 (E - E_x)^2/2}\cos{\left[ (E - E_x)\tau \right]},
     \label{eq:ge}
\end{eqnarray}

We establish Eq. (\ref{eq:ge}) as the reference for the data set structure and the subsequent analysis presented in sections \ref{sec:dataset}, \ref{sec:results-sim} and \ref{sec:results-dev}. When the tuning energy is equal to $E_x$, $-\overline{h}(E,\psi_I)$ provides the probability $P(x)$ to project the initial state $\ket{\psi_I}$ onto the eigenstate $\ket{x}$:
\begin{equation}
P(x) = -\overline{h}(E=E_x,\psi_I),
\label{eq:px}
\end{equation}
agreeing with the result predicted by Eq. (\ref{eq:probtotal}).

In the next section, we present the general structure of the Zeeman model. 
\section{The Zeeman Model}
\label{sec:zeeman}

In this section, we show how to apply the Rodeo Algorithm and explore the simulation outcomes for spin interactions described by the Zeeman model, which can be mathematically expressed by the following Hamiltonian:
\begin{eqnarray}
H_{\text{obj}} = -\sum_{i=1}^{M} \vec{\mu}_i\cdot\vec{B} =  -\mu_B\sum_{i=1}^{M} \left(\frac{\vec{L}_i}{\hbar} +\vec{\sigma}_i\right)\cdot\vec{B},
\label{eq:zeeman}
\end{eqnarray}
where $\vec{B}$ represents the magnetic field, $\mu_B = {e \hbar}/{2 m_e c}$ is the Bohr magneton and the operators $\vec{L}_i$ and $\vec{\sigma}_i$
are respectively the angular momentum and the Pauli spin matrices that act on the $i$-th particle. Additionally, $c$ and $\hbar=h/2\pi$ represent the speed of light and the reduced Planck constant, while $e$ and $m_e$ denote respectively the electron charge and its mass.  

For the sake of simplicity, we focus on the simplest case where $\vec{L}_i=0$ and set the units reduced such that $\mu_B=1$. Furthermore, without loss of generality, we consider that the magnetic field direction is aligned on the $\hat{z}$ axis, i.e. $\hat{z} \equiv \vec{B}/|B|$. Therefore, Eq. (\ref{eq:zeeman}) is reduced to
\begin{eqnarray}
H_{\text{obj}} = - B \sum_{i=1}^{M}\sigma_{z}.
    \label{eq:ZemmanHam}
\end{eqnarray}
Note that $H_{\text{obj}}$ does not entangles the system qubits that characterize $\ket{\psi_I}$, since the respective magnetic fields acts separately on each spin.  

The data set and the methodology addressed in the algorithms are detailed in the following subsection.
\section{Data Set Structure and Methodology}
\label{sec:dataset}

Applying Rodeo Algorithm to a given system can generate a large amount of data even for the simplest cases. As a typical Monte Carlo experiment (for which an average of the desired quantities can be obtained over a repeated random sampling~\cite{landau2021book}), the Rodeo scheme must be performed in a suitable number of times to reach precise estimation for the Schr\"odinger equation solutions. Therefore, the sampling process requires significant time and computational resources that may not always be readily available.

In this context, the storage method employed in this work utilizes a time series scheme, which registers the algorithm inputs and outputs for each shot. The total number of repetitions is represented by $N_{\text{rides}}$, and the data set structure is organized as follows: two $N_{\text{rides}}$-order arrays $\mathbf{X}$ and $\mathbf{Y}$ are first established such that its $i^{th}$ components are respectively given by
\begin{eqnarray}
X_i & = &  [\zeta_0,\zeta_1,...,\zeta_{N-1} ]\label{eq:data} \ \text{and}\nonumber\\
Y_i & = &  [t_0,t_1,...,t_{N-1},E_i,d,\tau,\{\omega\},\ket{\psi_I}].
\label{eq:X_iY_i}
\end{eqnarray}

The $X_i$ variables represent the measurement outcomes of the $k^{th}$ ancilla performed at the final stage of the Rodeo Algorithm, which is defined specifically for each platform. While in Pennylane $\zeta_k$ it is a float number that stores the expected value of $\sigma_z$, i.e. \ $\zeta_k = \langle \sigma_z \rangle_{k}$, in Qiskit it describes two integers $N_{\uparrow}$ and $N_{\downarrow}$ such that $\langle \sigma_z \rangle_{k} = (N_{\uparrow} - N_{\downarrow})/(N_{\uparrow} + N_{\downarrow})$.

In turn, the first $N$ elements of $Y_i$ correspond to the random time interval generated through the PDF for each ancilla. The array also stores the guessed value for the eigenvalue $E$, as well as the PDF parameters $d$ and $\tau$. Moreover, a set of parameters $\{\omega\}$ required to establish the Hamiltonian $H_{\text{obj}}$ is subsequently included on the list, which is followed by a bitstring that defines the initial state $\ket{\psi_I}$. Thereby, the data is structured as
\begin{eqnarray}
\text{data set} & = &  \{(X_0,Y_0),\cdots,(X_{N_{\text{rides}}-1},Y_{N_{\text{rides}}-1})\}.
\label{eq:dataset}
\end{eqnarray}

Once $Y_i$ is defined, we start the code by inferring  $\ket{\psi_I}$ with arbitrary coordinates through the Bloch sphere. On the sequence, the energy range is limited from $E_0$ to $E_f$ and divided into intervals of $\Delta E$, generating $N_E=\left({E_f-E_0}\right)/\Delta E$ levels.
The initial state is then fixed as a reference for a total of $N_{E} \times N_{\text{rounds}}$ repetitions of the Rodeo Algorithm, wherein $N_{\text{rounds}}$ cycles are performed at each energy level to store the outputs in $X_i$. Finally, we set another guess for $\ket{\psi_I}$ and repeat the scheme $N_{\psi}$ times, which results in a total of $N_{\text{rides}}= N_{E} \times N_{\text{rounds}} \times N_{\psi}$ steps.

Due to the fact that real quantum computers offer access to a limited number of qubits in open access accounts, it is necessary to establish a balance between the system size and the number of ancilla available in the circuit. Since the solution accuracy is inversely related to the latter on real devices, we decide to conduct the iterations in Pennylane and Qiskit prioritizing the minimum possible value to reduce this dependence; i.e. limiting the range in Eq. (\ref{eq:psi0}) to one ancilla ($N=1$). 

Finally, to foster a collaborative environment and facilitate further scientific studies through database analysis, we also want to emphasize that the data sets and computational codes developed here are available on Zenodo \cite{ZenodoDataPublication}.

In the next sections, we apply the mathematical and computational tools presented in sections \ref{sec:bullOper}, \ref{sec:zeeman}, and \ref{sec:dataset} to develop the Rodeo Algorithm for Hamiltonians described by the Zeeman model in systems with one and two spins. 
\section{Results Obtained in the Simulator}
\label{sec:results-sim}
\subsection{ One-Spin Hamiltonian}
\label{subsec:onespin}

We initialize this study by considering a system consisting of a single atom in Eq. (\ref{eq:ZemmanHam}), i.e. $M=1$. Hence, the Hamiltonian can be represented as follows:
\begin{eqnarray}
H_{\text{obj}} = - B\sigma_{z}.
\label{eq:onespin}
\end{eqnarray}
This model serves as a suitable prototype to validate the Rodeo Algorithm since the solution to the time-independent Schr\"odinger equation is straightforward. The eigenstates of $H_{\text{obj}}$ are the computational basis states $\ket{0}$ and $\ket{1}$ themselves, with the respectively eigenvalues given by $E_{0} = -B$ and $E_{1} = +B$.

In turn, the target system is initially considered in an arbitrary state as
\begin{equation}
\ket{\psi_I} = \cos{\left(\frac{\theta}{2}\right)}\ket{0}  +  e^{i\varphi}\sin{\left(\frac{\theta}{2}\right)}\ket{1}, 
\label{eq:SN1G}
\end{equation}
where $\theta$ and $\varphi$ are respectively the polar and azimuthal angles in the Bloch sphere. It is worth noting that the latter imposes a phase that does not affect the measurement process. Thus, without loss of generality, we can set $\varphi = 0$ and describe $\ket{\psi_I}$ solely using the polar angle $\theta$. In this case, applying Eq. (\ref{eq:SN1G}) on Eq. (\ref{eq:ge}) leads to
\begin{align}
\overline{h}(E,\psi_I) &=-\sin^{2}{\left(\frac{\theta}{2}\right)}e^{-[d(E+B)]^2/2}\cos{[\tau(E+B)]}  \nonumber\\ 
            &-\cos^{2}{\left(\frac{\theta}{2}\right)} e^{-[d(E-B)]^2/2} \cos{[\tau(E-B)]}.
\label{eq:getheta1}
\end{align}

According to the original proposal, the Rodeo Algorithm successfully detects the eigenvalues and eigenstates when $P(x)=|\bra{x}\ket{\psi_I}|^{2} \approx 1$.
Therefore, from Eqs. (\ref{eq:px}) and (\ref{eq:getheta1}), the condition $-\overline{h}(E,\psi_I)\approx 1$ is satisfied for each eigenvalue $E_{j}$ with $j \in \;\{0,1\}$ if 
\begin{eqnarray}
|\sin{(\theta/2)}|^2e^{-[d(E_{x}+B)]^2/2}\cos{[\tau(E_{x}+B)]} \approx 1 \;\;\text{or}\nonumber\\
|\;cos{(\theta/2)}|^2 e^{-[d(E_{x}-B)]^2/2} \cos{[\tau(E_{x}-B)]} \approx 1,
\label{eq:overlapsigmaz}
\end{eqnarray}
where the solutions $\theta \approx \{\pi;0\}$ are respectively associated to $E_{0}$ and $E_{1}$.

The data analysis and techniques proposed to detail and improve the Rodeo algorithm are discussed in the next topics.  
\subsubsection{Data Analysis
\label{subsubsec:data}}

Given the constraints of major NISQ devices and the scarcity of computational resources in a real quantum computer, we start to investigate the one-spin Zeeman model by exploiting the capabilities of the PennyLane platform {\cite{refpennylane}. The simulator developed by Xanadu is a differentiable programming framework for quantum computing, whose parameters optimization through gradient methods will be useful in future studies to explore the landscape of Eq. (\ref{eq:ge}) for other models. 

The results for the scenario presented in Eq. (\ref{eq:getheta1}) for $N=1$ ancilla are plotted against $\theta/\pi$ and $E$ in Fig.~\ref{fig:gETheta}, where $-\overline{h}(E,\psi_I)$ corresponds to the numerical estimation of
\begin{equation}
-\overline{h}(E,\psi_I) = \frac{1}{N_{Rounds}}\sum_{i=1}^{N_{Rounds}}-h(E,\psi_I|t_i),
\label{eq:MedArit}
\end{equation}
considering that $h(E,\psi_I|t_i)$ is given by Eq. (\ref{eq:h_psi_f}) and $N_{\text{rounds}}=50$. The color scheme ranges from red ($\theta=0$) to yellow ($\theta=\pi$), and the values for $t_{N}$ were generated a hundred times in a normal distribution for each value of $E$, with $B=1.0$, $\tau=10$ and $d=7$. Note that the exact solutions represented by the continuous solid black lines confirm the predictions of Eq. (\ref{eq:px}), where it was pointed out that $P(x) = -\overline{h}(E=E_x,\psi_I)$. 
\begin{figure}[!ht]
\includegraphics[scale=0.41]{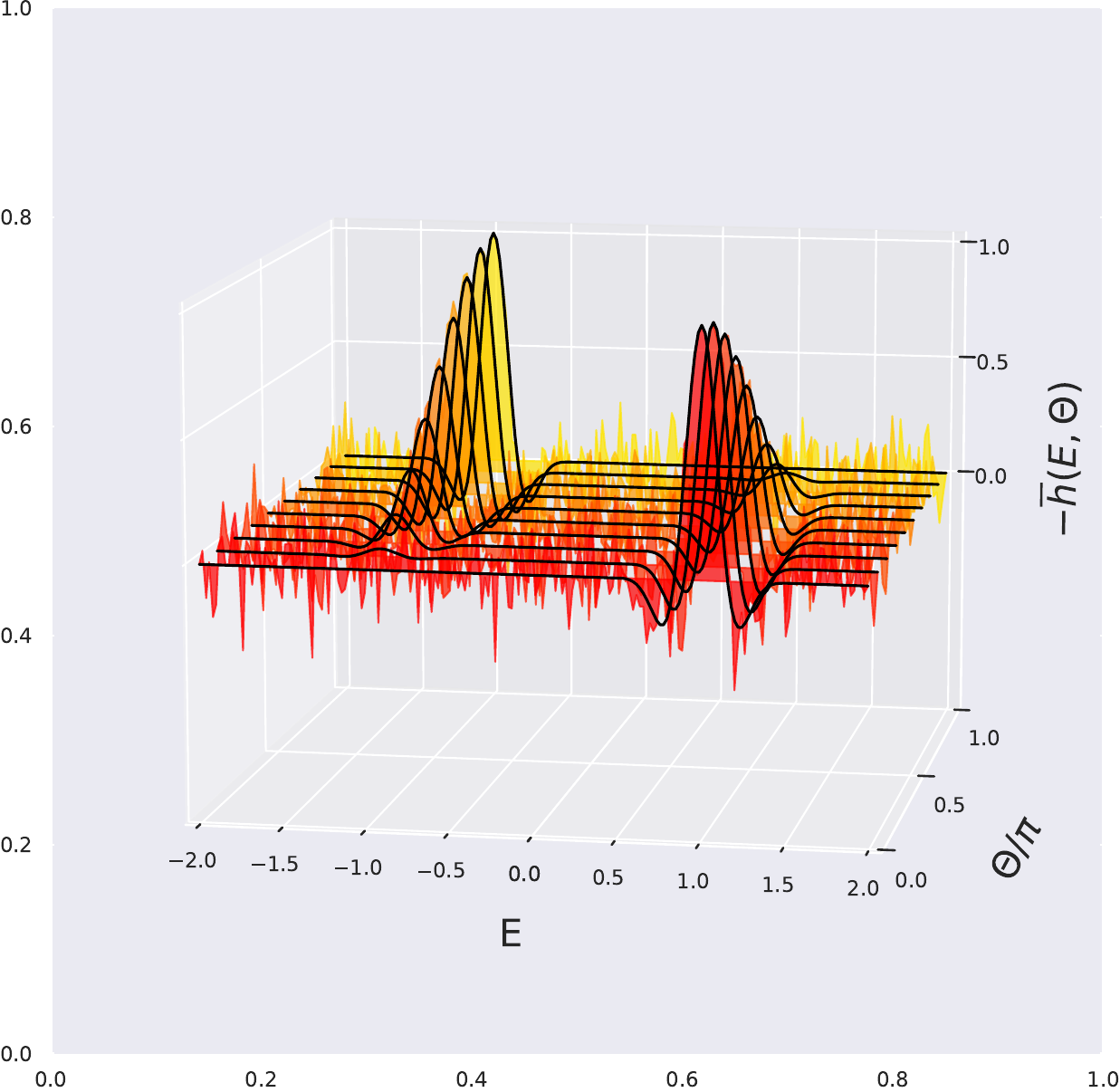} 
\caption{Results for the circuit build on Pennylane} for the one-spin Zeeman model with $N=1$ ancilla, considering $B=1.0$, $\tau=10$, $d=7$ and $N_{\text{rounds}}=50$.
\label{fig:gETheta}
\end{figure}
\begin{figure}[!ht]
    \includegraphics[scale=0.28]{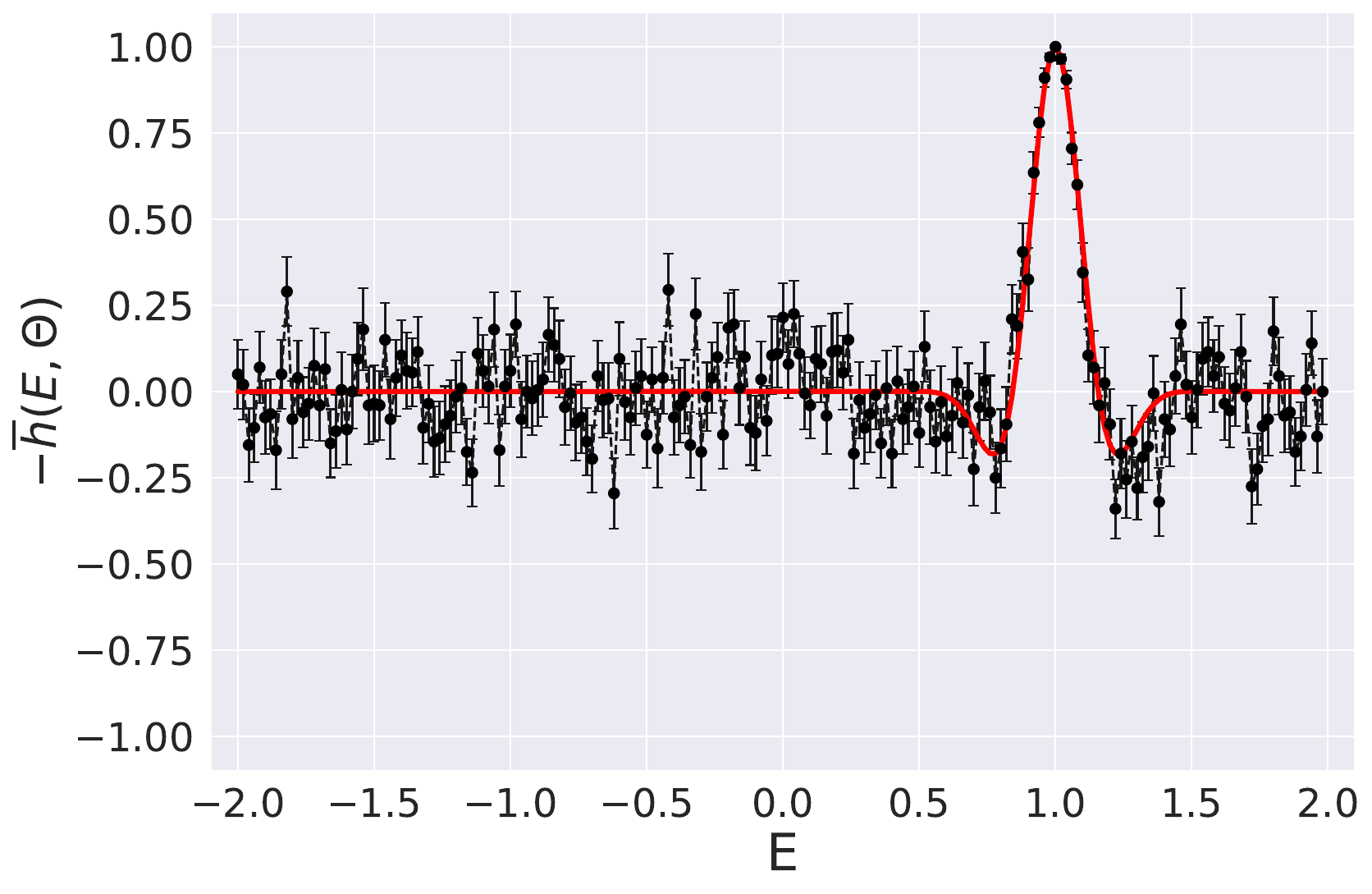} \\
    \includegraphics[scale=0.28]{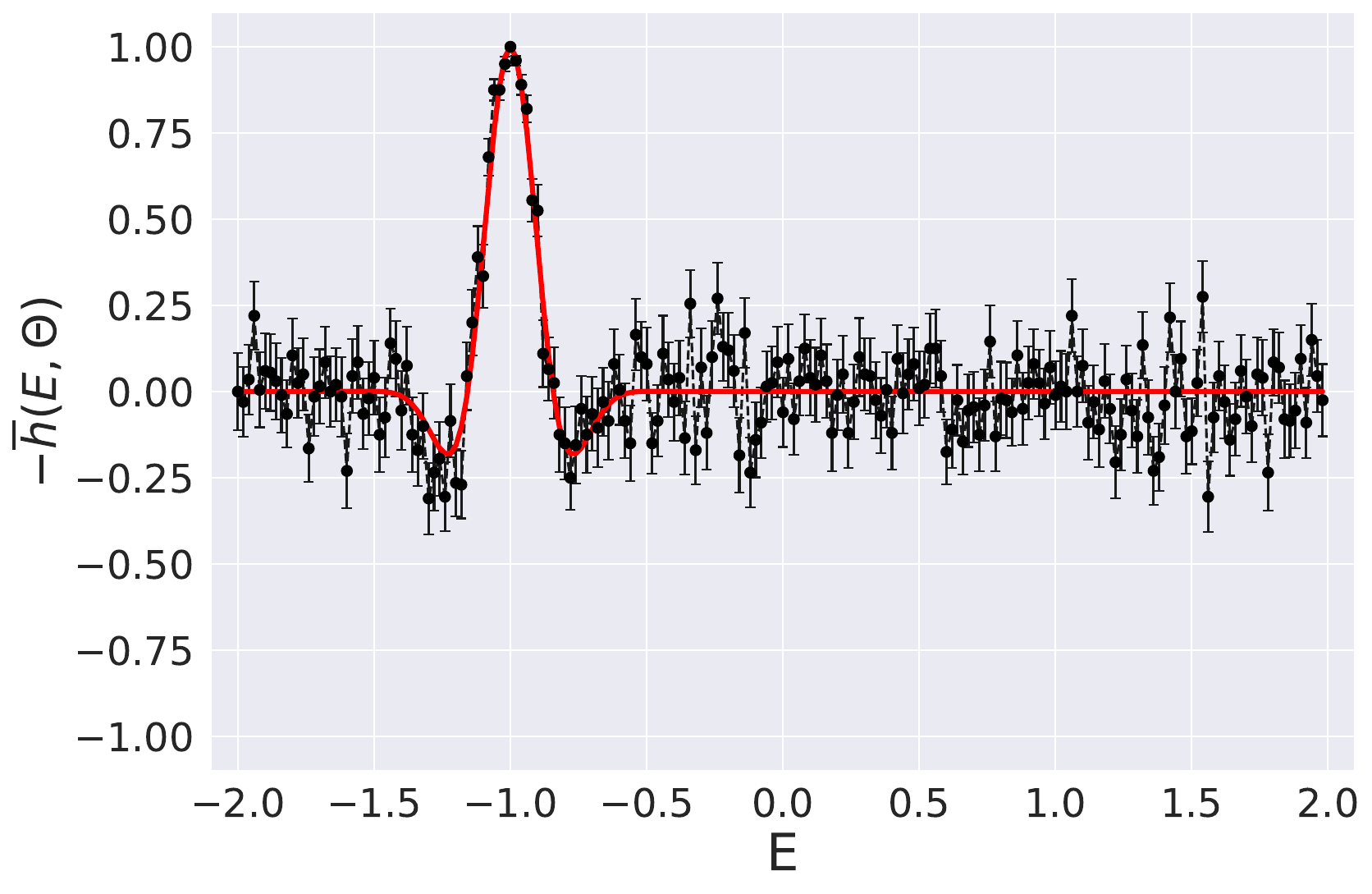}\\
    \includegraphics[scale=0.28]{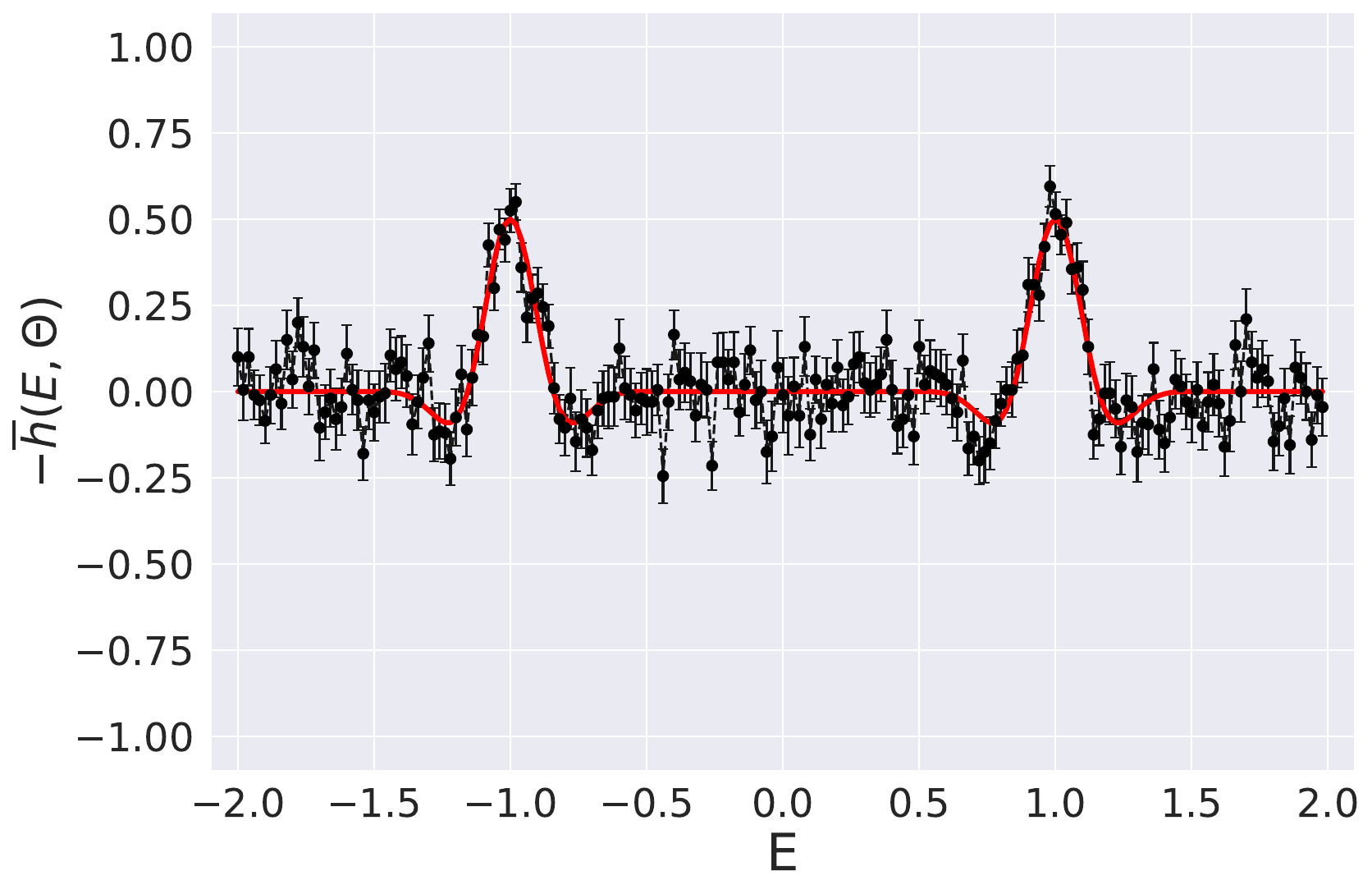} \\
    \includegraphics[scale=0.28]{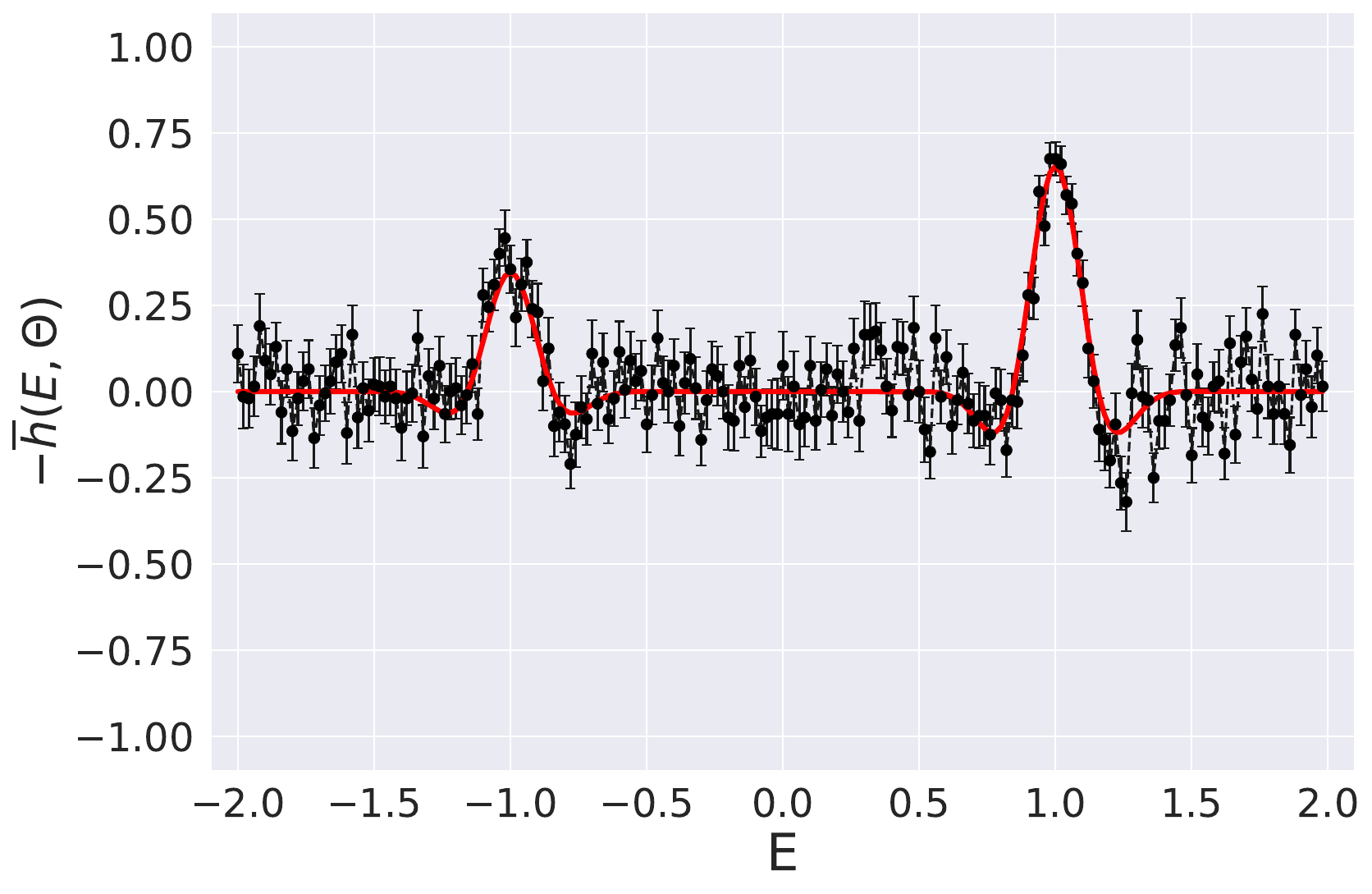}
    \caption{From top to down, specific results for the circuit build on Pennylane of the one-spin Zeeman model with $N=1$, $B=1.0$, $\tau=10$, $d=7$ and $N_{\text{rounds}}=50$ for $\theta= 
\{0; \pi; \pi/2; \pi/5\}$.}
\label{fig:gE1}
\end{figure}

This analysis can be more easily assessed if we look at specific values of $\theta$. To this end, Fig.~\ref{fig:gE1} displays slices at particular choices of $\theta$ for the surface presented in Fig.~\ref{fig:gETheta}. The solid red curve illustrates the exact results given by Eq.(\ref{eq:getheta1}), while the black dot symbols correspond to the solution of Rodeo Algorithm obtained in the simulator. In turn, the error bars represent the standard deviation given by 
\begin{equation}
s(h(E,\psi_I)) = \sqrt{\frac{\mathrm{Var}(h(E,\psi_I))}{N_{\text{rounds}}}},
\label{eq:sde}
\end{equation}
where $\mathrm{Var}(x) = \overline{(x^2)}-(\overline{x})^2$. We discuss the error analysis in more detail in the next subsection.

Counting from top to down, the first and second panels display the results for $\theta=0$ and $\theta=\pi$, which correspond respectively to the eigenstates $\ket{1}$ and $\ket{0}$. For both cases, all ancillary qubits collapse to the state $\ket{1}^{\otimes N}$ after its measurement. In this sense, the functions achieve the maximum value allowed in a single peak for $-\overline{h}(E=+1.0,\theta=0)=1$ and $-\overline{h}(E=-1.0,\theta=\pi)=1$, confirming the full overlap between the initial state and the associated eigenstate.

However, contrary to the statements presented in the original proposal \cite{choi2021}, the function $-\overline{h}(E,\psi_I)$ still provides enough information to characterize $H_{\text{obj}}$ when the overlap is significantly less than one. The third panel of Fig.~\ref{fig:gE1} exhibit the results for $\theta=\pi/2$, which describes the state $\ket{+}=\ket{0}+\ket{1})/\sqrt{2}$. Since the initial state plane is symmetric to the eigenstate axes, two maxima points are observed at $E=\pm1.0$ such that $-\overline{h}(E = \pm 1.0,\theta=\pi/2)\approx 0.5$, indicating that both basis states are equally likely to be measured.

Finally, the last panel shows the results for $\theta =\pi/5$, where the expected probabilities of measuring the system in the states $\ket{1}$ and $\ket{0}$ are given respectively by $P(1) = \cos^2{(\pi/5)}$ and $P(0)=\sin^2{(\pi/5)}$. In this case, two peaks are observed at $E=\pm1.0$, where $P(1) \approx -\overline{h}(E=+1.0,\theta= \pi/5) \approx 0.66$ and $P(0) \approx -\overline{h}(E=-1.0,\theta=\pi/5) \approx 0.35$ satisfies the condition $P(0)+ P(1)\approx1$.

The numerical analysis developed for this example can be generalized to confirm the analytical claim that the algorithm provides the eigenstates and eigenvalues spectrum of $H_{\text{obj}}$ with arbitrary input. 
On one hand, the PDF will never reach its maximum value if the fidelity regarding  $\ket{\psi_I}$ and the Hamiltonian eigenstates is not close to one, since a significant portion of the ancillary measurements will result in $|0\rangle$ (due to the simultaneous contribution of the factors $e^{i\;E_{0}t}$ and $e^{i\;E_{1}t}$ in Eq. (\ref{eq:getheta1})). 
On the other hand, this scenario brings a trade-off: if there is no prior knowledge about $\ket{\psi_I}$, one can infer both eigenvalues at the same time through the pattern exhibited by the PDF, despite having a little uncertainty about $E_x$. 

In turn, the probabilities   $0 \leq P(x)=|\bra{x}\ket{\psi_I}|^2\leq 1$ given by the peaks amplitude can be used to estimate $\bra{\psi_I}|H_{\text{obj}}|\ket{\psi_I}$, as long as the condition $\sum_{x}{P(x)\approx1}$ is satisfied. Moreover, the slope of $-\overline {h}(E,\psi_I)$ can be manipulated to match the eigenstates. If the curve shows two peaks whose amplitude is less than one, we can modify the initial state parameters and reset the circuit. These changes will cause one of the peaks to increase, while the other decreases. If we continue to update the values of $\theta$ and $\phi$ in Eq. (\ref{eq:SN1G}) to enhance the gap, the lower peak is going to vanish whereas the other will reach its maximum value for a given $E_{x}$ - at this point, $\ket{\psi_I}$ is the eigenstate itself. 
Consequently, this method can be applied to detect any eigenstate $\ket{x}$ and characterize the Hamiltonian spectrum along with the respective eigenvalues.

Through the next subsections, we discuss four directions to follow to diminish the floating points of $-\overline{h}(E,\psi_I)$ that differ from the wanted eigenvalues near its peak. These strategies rely on repeating the measurements for other sets of the $t_{N}$ distribution, increasing the number of auxiliary qubits available to the system, defining a suitable value for the standard deviation, and optimizing the initial state settings.
\subsubsection{Measurement repetition}
\label{subsubsec:measrep}

\begin{figure}[!ht]
    \includegraphics[scale=0.15]{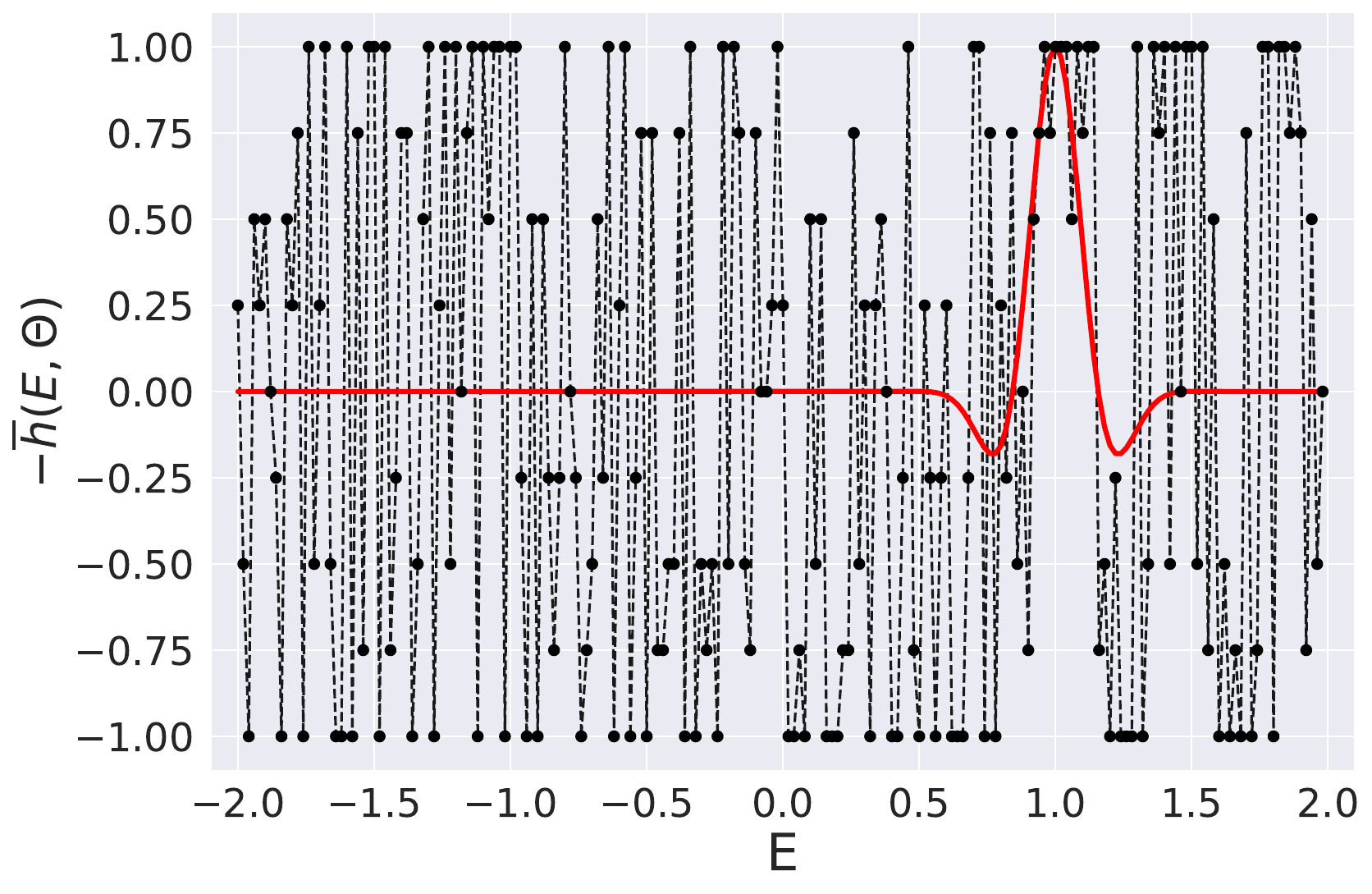} 
    \includegraphics[scale=0.15]{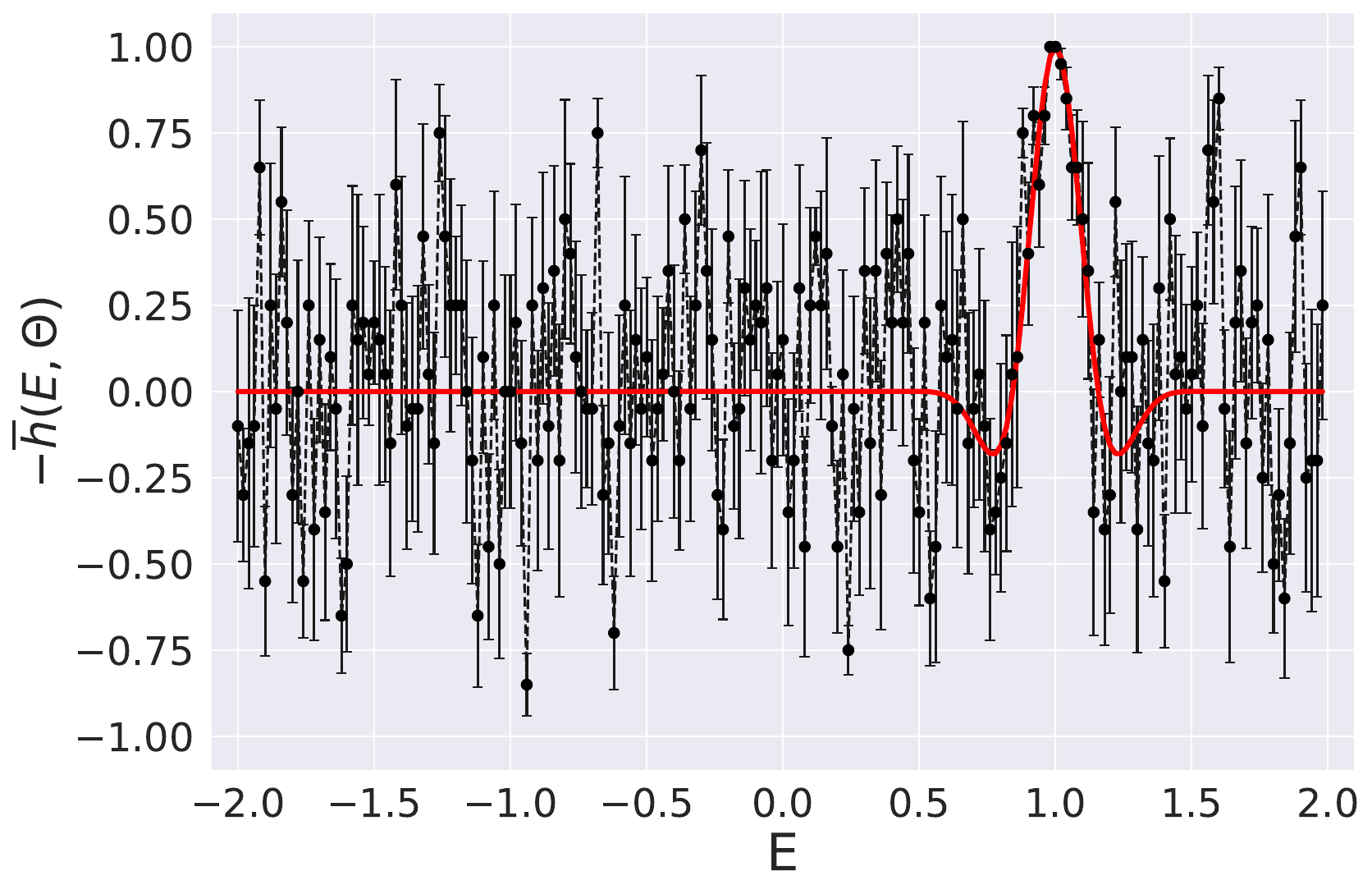} \\
    \includegraphics[scale=0.15]{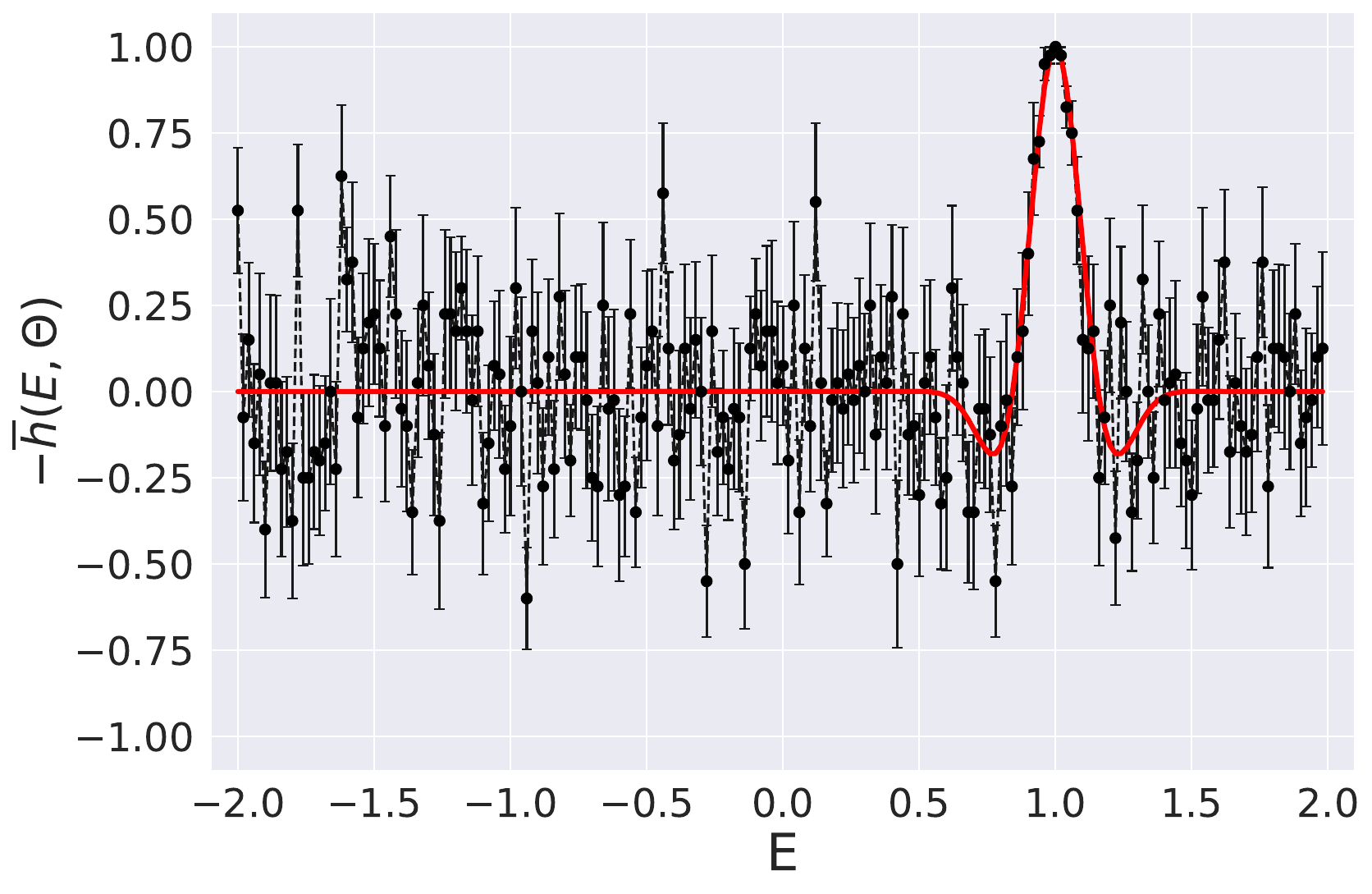} 
    \includegraphics[scale=0.15]{ZemmanBz+-06ThetaNmed50_d7_tau10/Zemman2Dimages/Zemman_RXequalPiover2RYequalPiover30-Bx=By=0_Bz=1.00_theta=0.00_Nmed_50.pdf}     
   \caption{From top to down, specific results for the circuit build on Pennylane of the one-spin Zeeman model with $N=1$, $B=1.0$, $\Theta=\pi/30$, $\tau=10$ and $d=7$ for $N_{\text{rounds}}=\{1; 5; 10; 50\}$.}
\label{fig:plszn5e-020med10mod}
\end{figure}

Repeating an experiment under similar conditions is essential to understand any phenomenon and assess the uncertainty associated with the methodology adopted throughout the process. In statistical analysis, this uncertainty is estimated by the dispersion, which can be quantified by the standard deviation related to the spread of experimental results around their mean value. Based on this principle, we can repeat the measurement process $N_{\text{rounds}}$ for a fixed initial state $|\psi_{I}\rangle$ to reduce the dispersion. This strategy consists of choosing random values of $t_{N}$ for each $E \in [E-\epsilon, E+\epsilon]$ with $\epsilon \approx 0$, and record the output given by Eq. (\ref{eq:MedArit}) to evaluate another average over the expected value. 

Fig. \ref{fig:plszn5e-020med10mod} shows the results for a circuit with $N=1$ and $\theta=\pi/30\approx 0.1$ for $N_{\text{rounds}}=\{1; 5; 10; 50\}$, where the other parameters remain unchanged. With the appropriate choice of repetitions, the method dismisses the need to expand the number of ancilla and is highly effective for circuits with few qubits. Note that the unavoidable fluctuation is considerably reduced as the parameter increases, where no points but the eigenvalues reach $-\overline{h}(E,\psi_I)=1$ for $N_{\text{rounds}}=50$ (as in Fig. \ref{fig:gE1}). 

However, even in the absence of noise, systems with only one ancilla will collapse to $|1\rangle$ for every value of $E=E_{\text{obj}}-2j\pi$ with $j\in \mathbb{N}$ in Eq. (\ref{eq:ge}), and the measurement process may not be sufficient to guarantee the filtering of eigenvalues. To avoid this problem, we can fix the distribution for $t_{N}$ and work in systems with more auxiliary qubits, as discussed in the next subsection.
\subsubsection{Increasing the number of ancillary qubits}
\label{subsubsec:increaseancilla}
\begin{figure}[!ht]
    \includegraphics[scale=0.15]{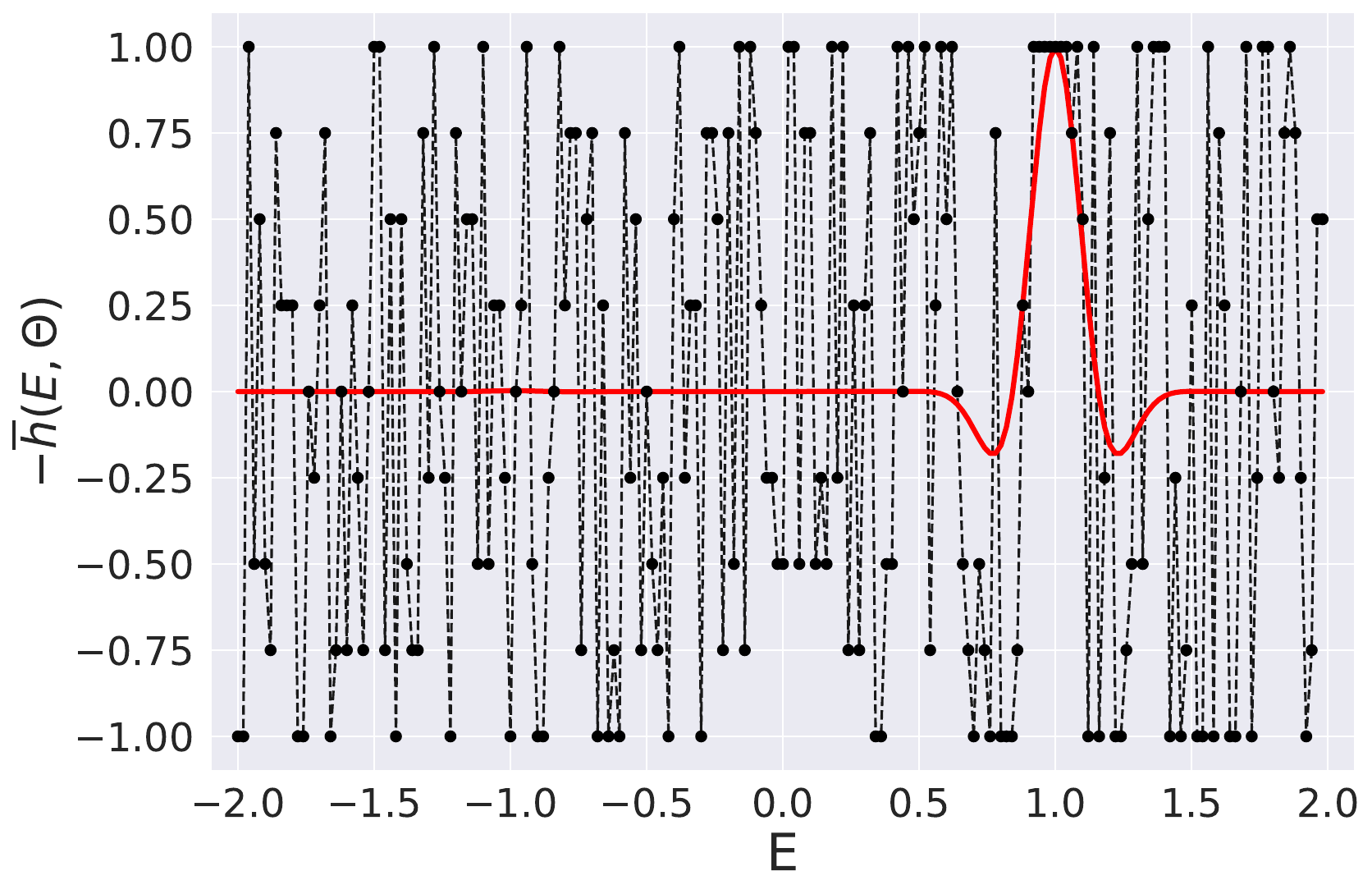}
    \includegraphics[scale=0.15]{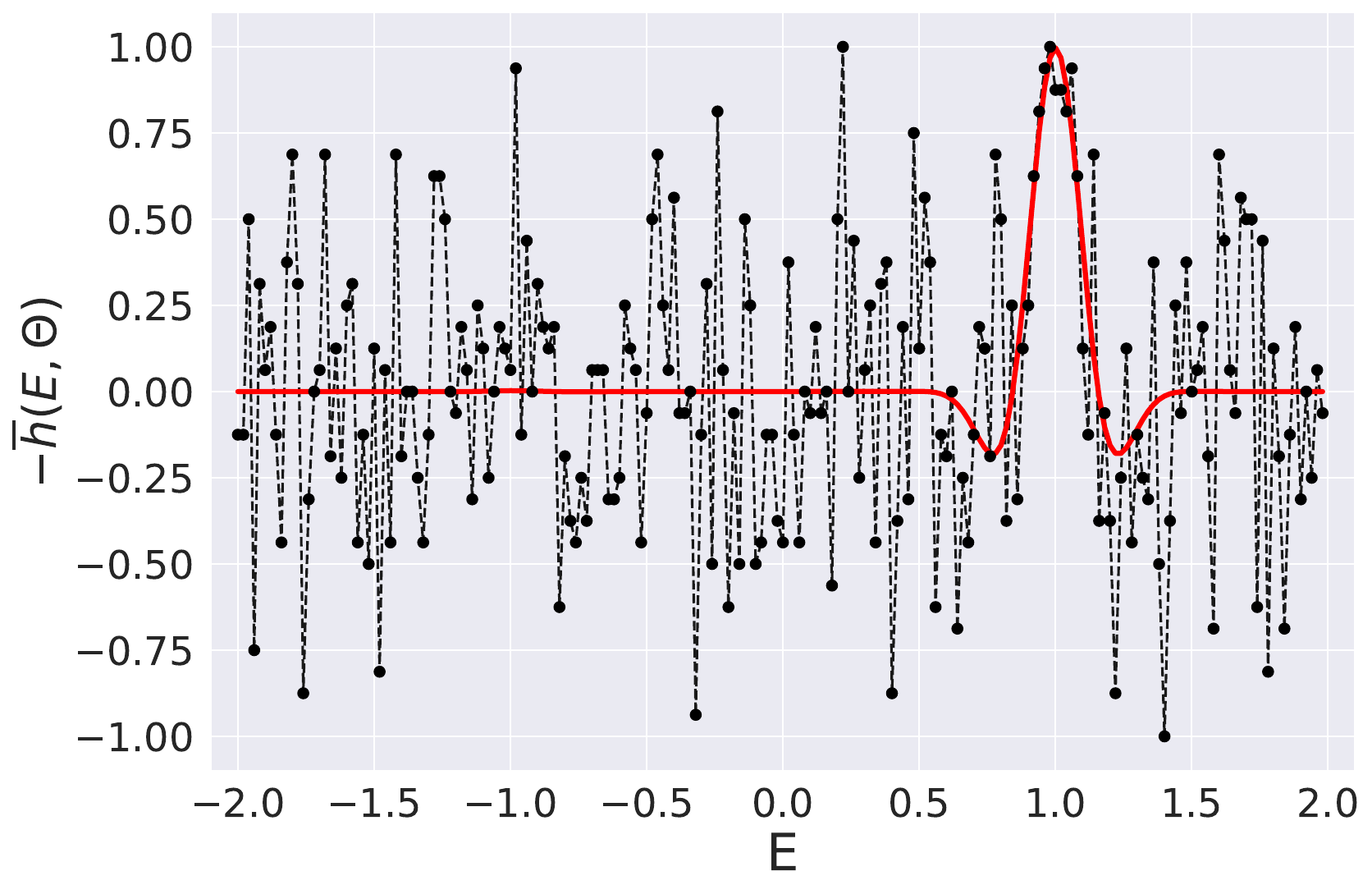}\\
    \includegraphics[scale=0.15]{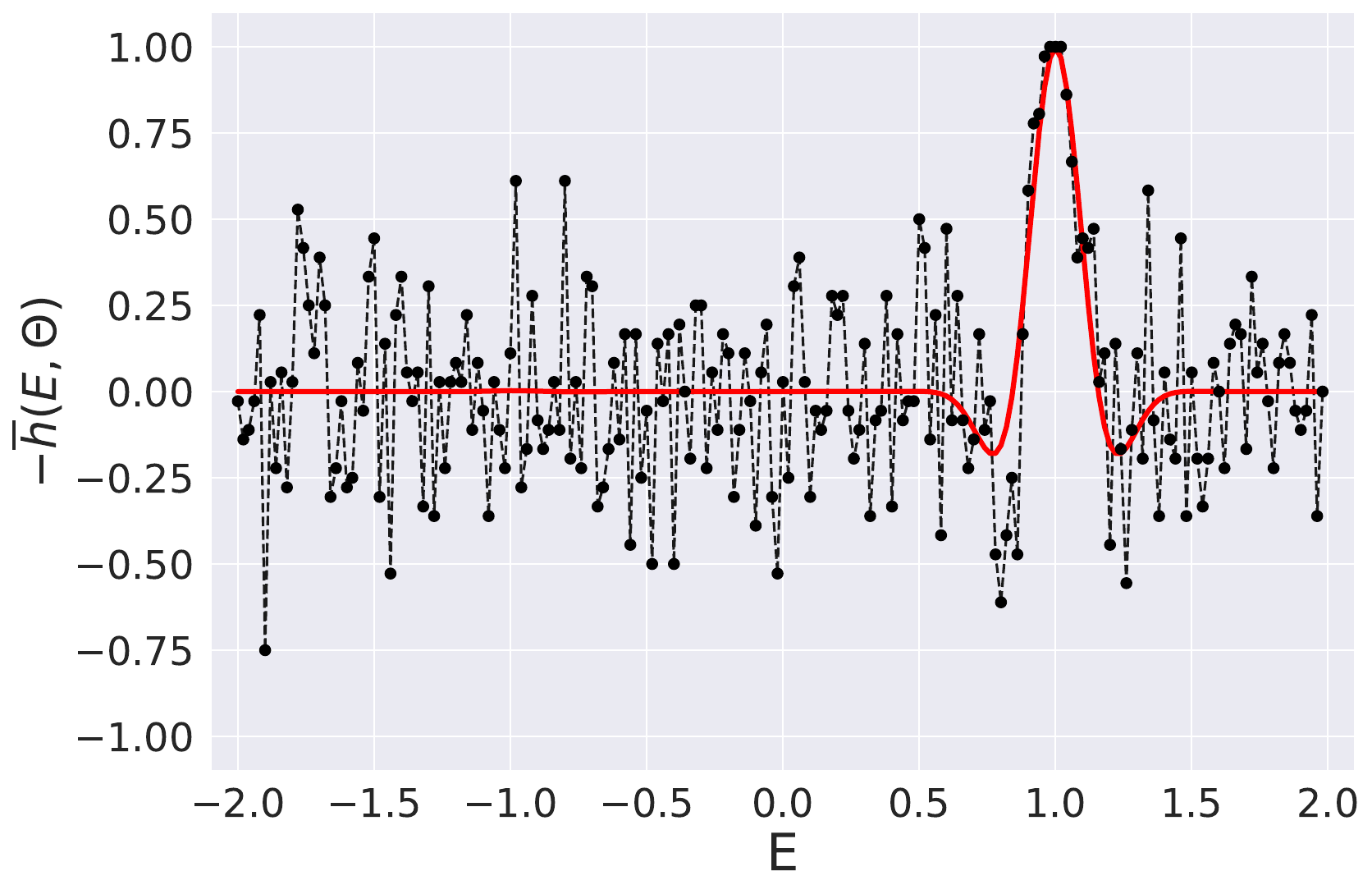}
    \includegraphics[scale=0.15]{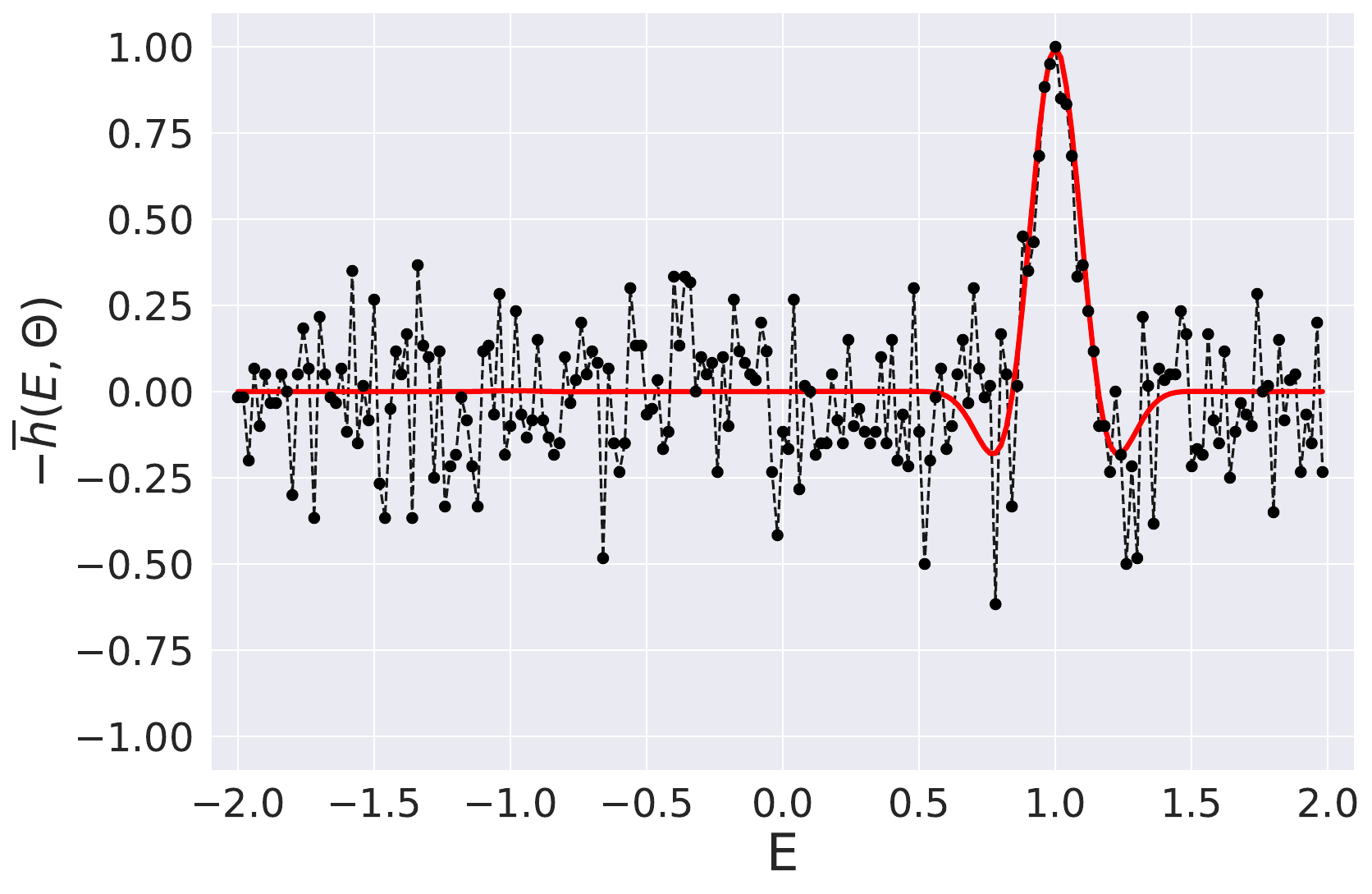} 
    \caption{From top to down, specific results for the circuit build on Pennylane of the one-spin Zeeman model with $B=1.0$, $\theta=\pi/30$, $\tau=10$, $d=7$ and $N_{\text{rounds}}=1$ for $N=\{1;4;9;15\}$.}
\label{fig:plszn5_10mod}
\end{figure}

The statistical sampling can also be improved by increasing the number of auxiliary qubits, which can be considered as a parallel processing of the algorithm on a single ancilla. Fig. \ref{fig:plszn5_10mod} shows the results for a circuit similar to Fig. \ref{fig:plszn5e-020med10mod} with $N_{\text{rounds}}=1$ for $N=\{1;4;9;15\}$. Note that as $N$ increases (despite a few isolated points with an approximate behavior), 
the data also that converges smoothly to $E=E_{\text{obj}}$ when $-\overline{h}(E,\psi_I)=1$, confirming that all ancilla were found at the $|1\rangle^{\otimes N}$ state after its measurement.

In fact, systems with more auxiliary qubits must provide better accuracy at the results, as the $t_{N}$ distribution for each Rodeo cycle expands the amount of data available at the circuit in one execution. At a certain point, there is no need to repeat the process, since a single measurement of the system for a fixed value of $E \in [E-\epsilon, E+\epsilon]$ is sufficient to give a reliable outcome.

On the other hand, as discussed in \cite{rocha2023}, each ancilla must be connected to three individual operations (due to the Hadamard and the Phase gates) and the controlled-gate that dictates the target system evolution. Therefore, adding qubits to the circuit increases its spatial and time complexity \cite{barenco1995, bernstein1997, bouland2019}, which are respectively associated to the circuit´s width (defined by the total number of qubits) and depth (associated with the time required to perform these operations). Hence, this choice makes the system more vulnerable to noise in real scenarios \cite{nielsen2010book}, since the decoherence process exerts a significant influence on NISQ devices \cite{refqiskit}.

In the next subsection, we show how the results are also influenced by modifying the PDF parameters.
\subsubsection{Tuning the parameters of the probability density function}
\label{subsubsec:tuningpdf}
\begin{figure}[!ht]
    \includegraphics[scale=0.15]{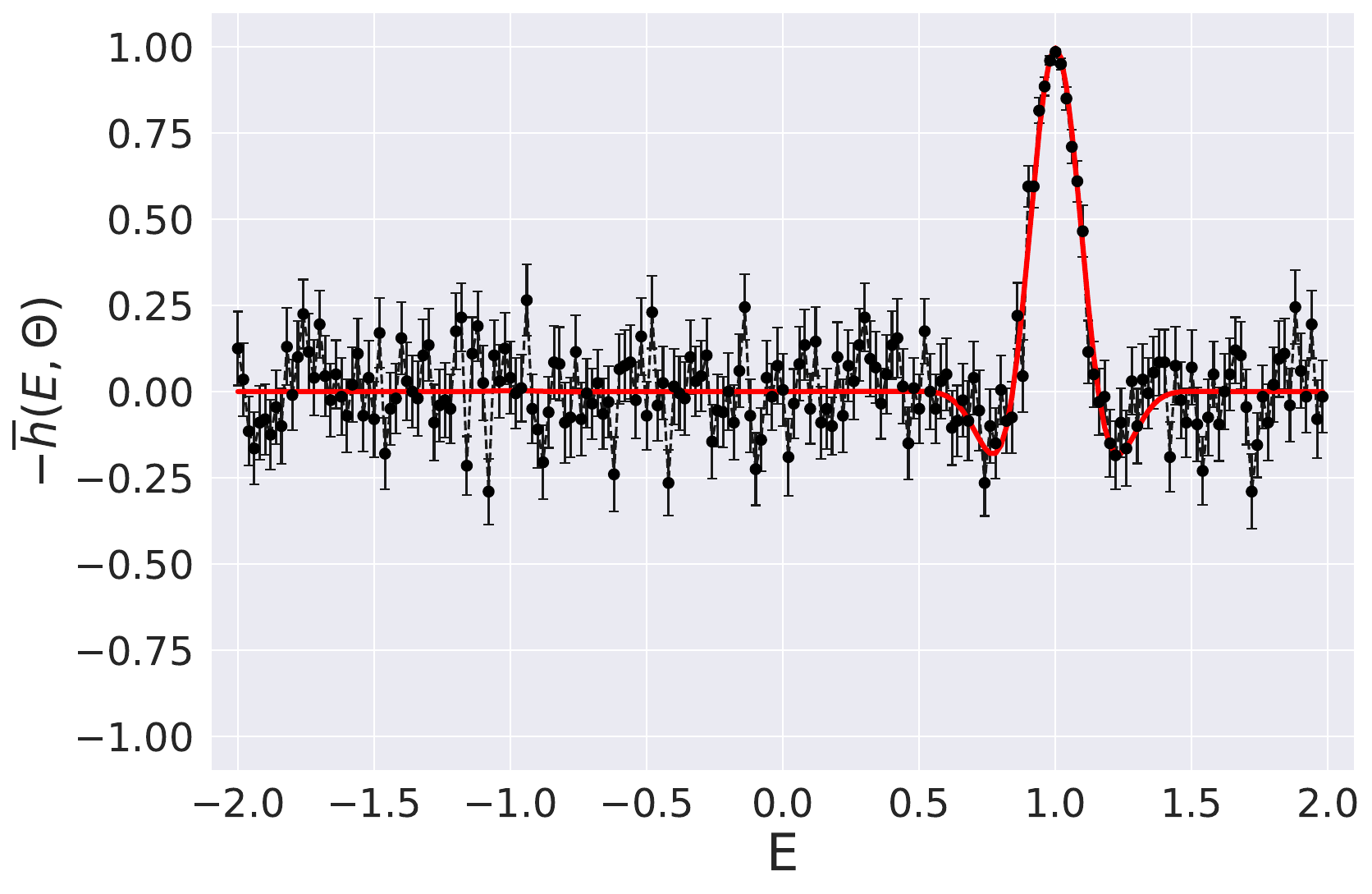}
    \includegraphics[scale=0.15]{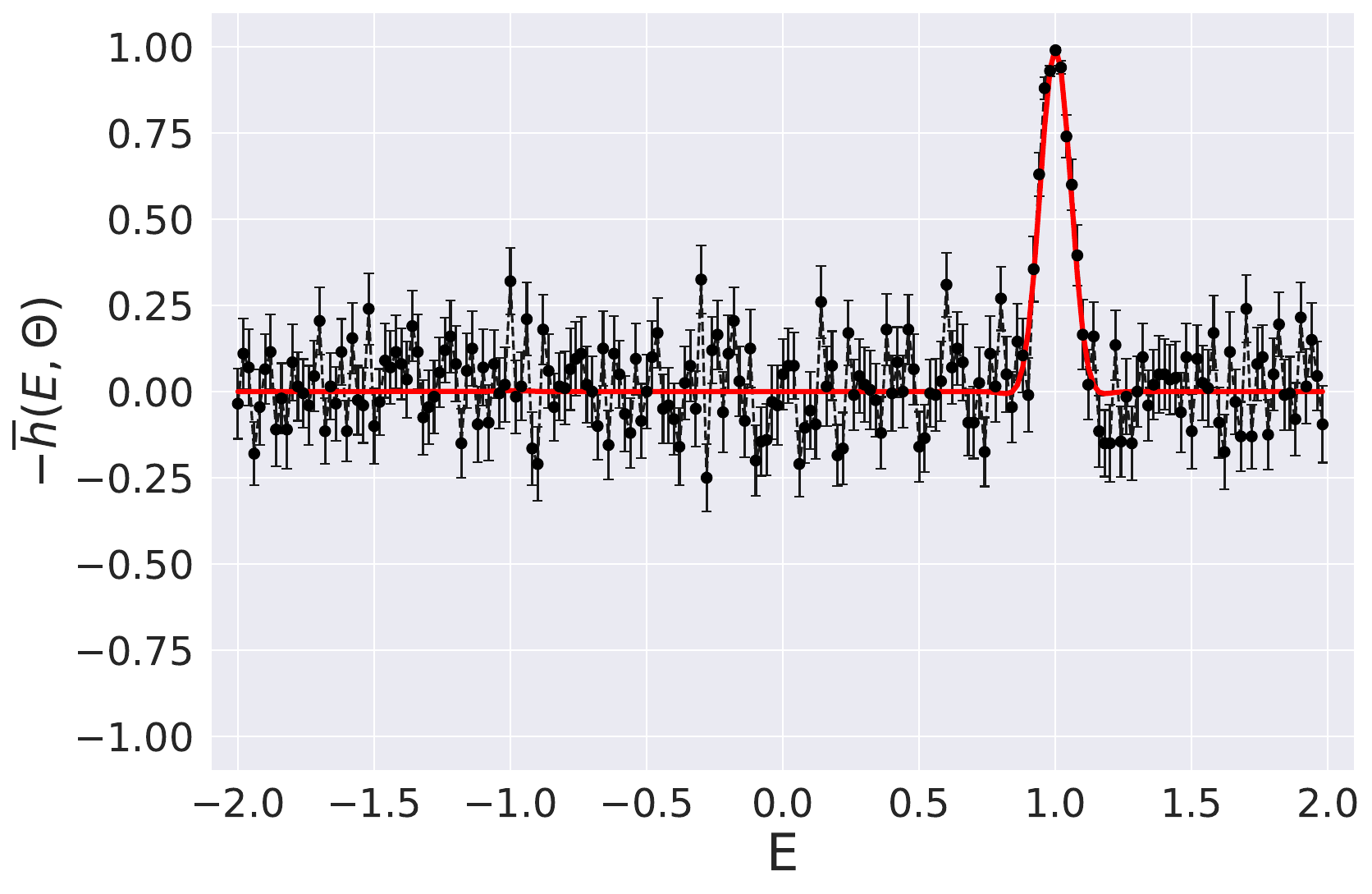}\\
    \includegraphics[scale=0.15]{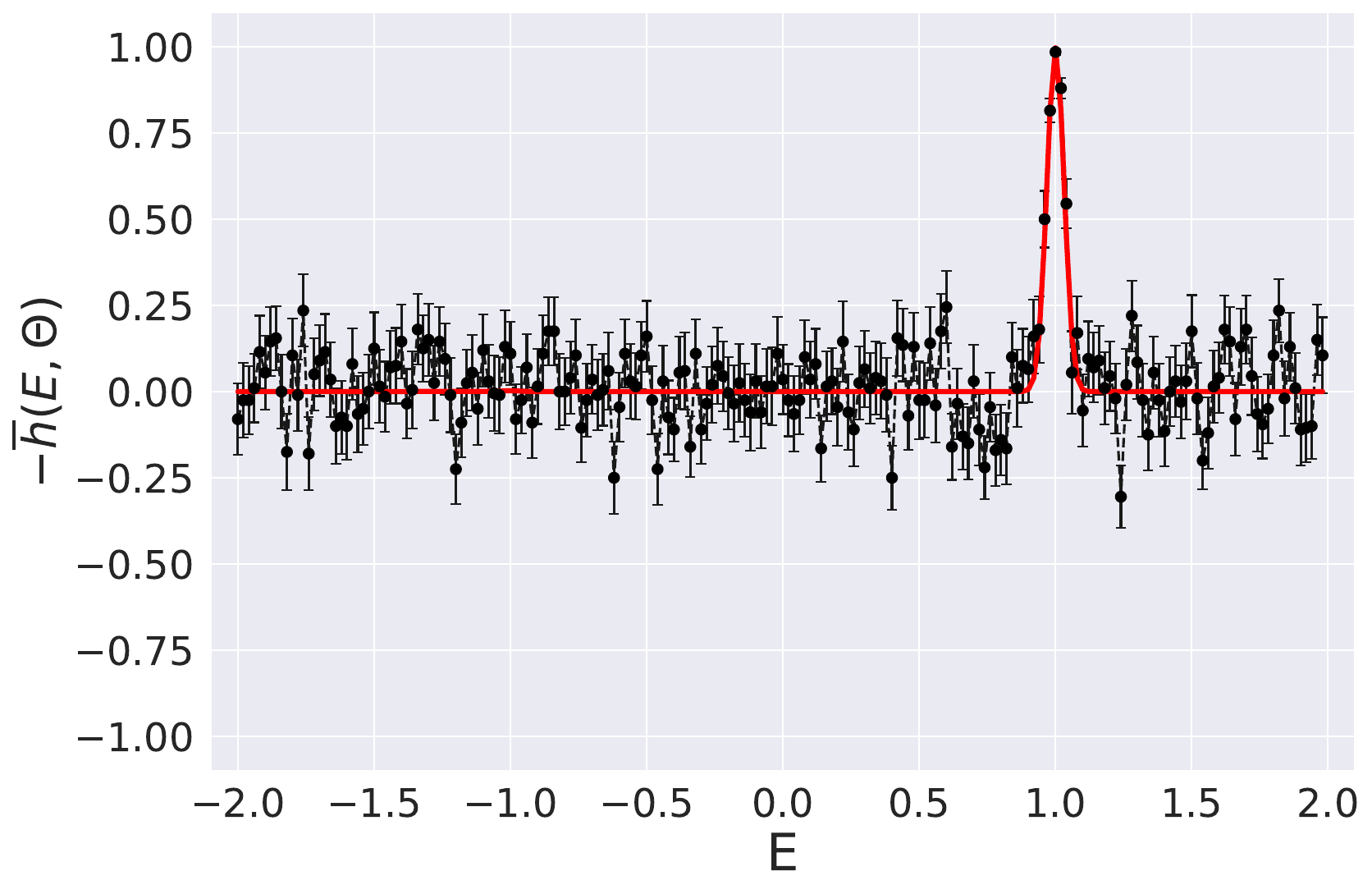}
    \includegraphics[scale=0.15]{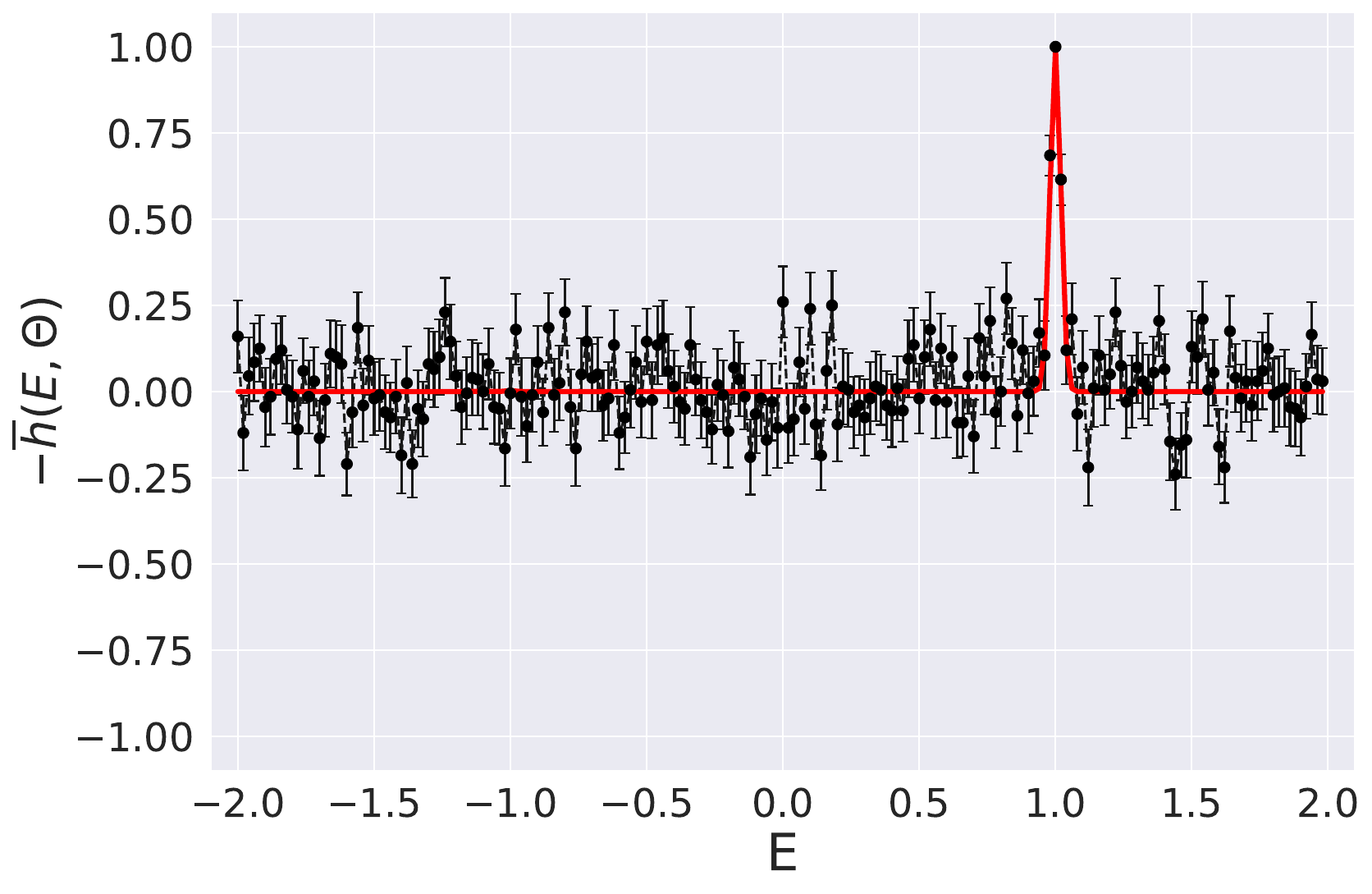}\\
    \caption{From top to down, specific results for the circuit build on Pennylane of the one-spin Zeeman model with $N=1$, $B=1.0$, $\theta=\pi/30$, $N_{\text{rounds}}=50$ and $\tau=10$ for $d=\{7;15;30;50\}$.}
\label{fig:plszn5e-020med10sigmod}
\end{figure}
\begin{figure}[!ht]
    \includegraphics[scale=0.15]{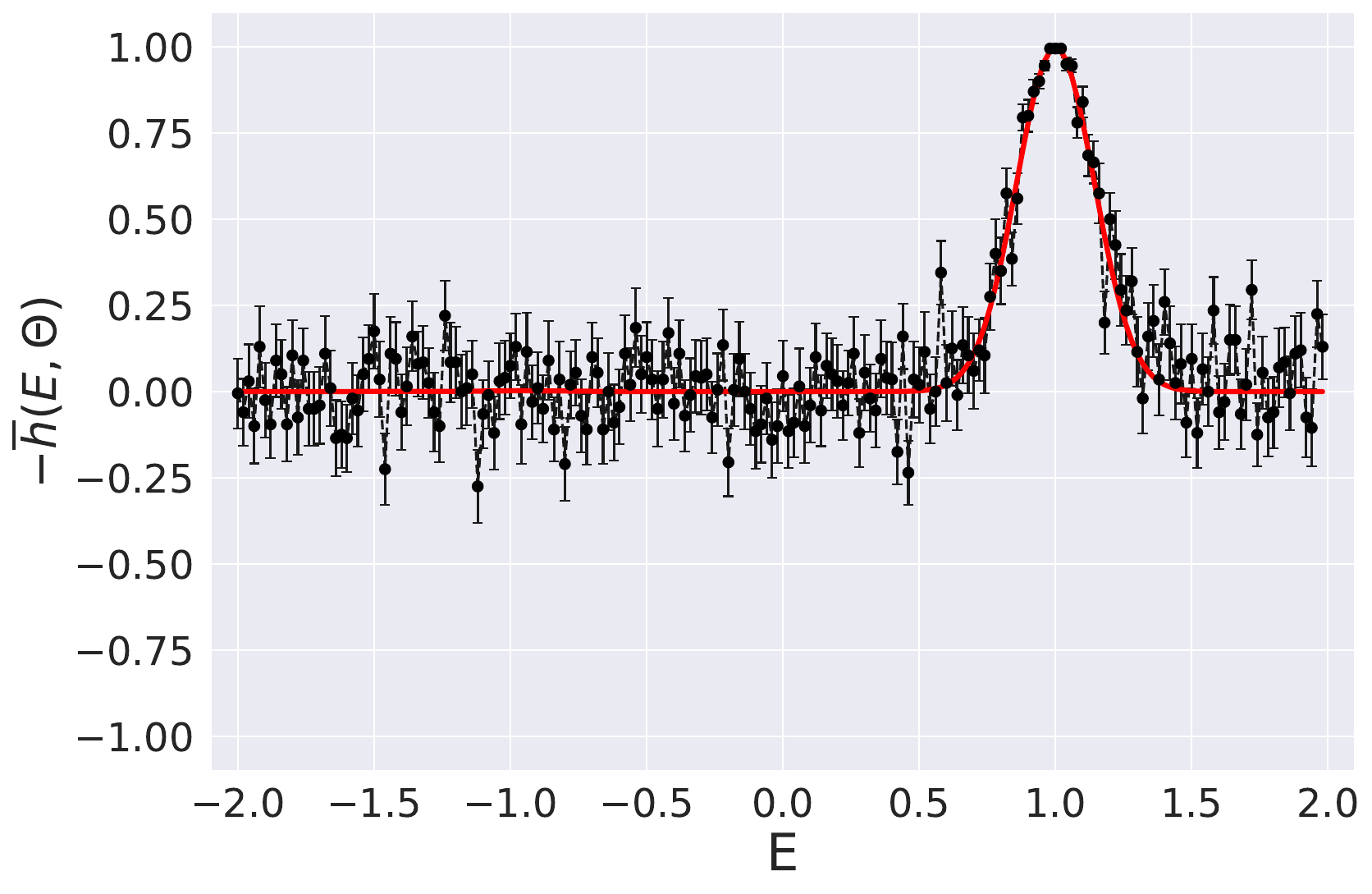}
    \includegraphics[scale=0.15]{ZemmanBz+-06ThetaNmed50_d7_tau10/Zemman2Dimages/Zemman_RX=RY_pi_over_30-Bx=By=0_Bz=1.00_theta=0.10_Nanc=1.00_tau10.0_d7.0_Nmed_50.pdf}\\
    \includegraphics[scale=0.15]{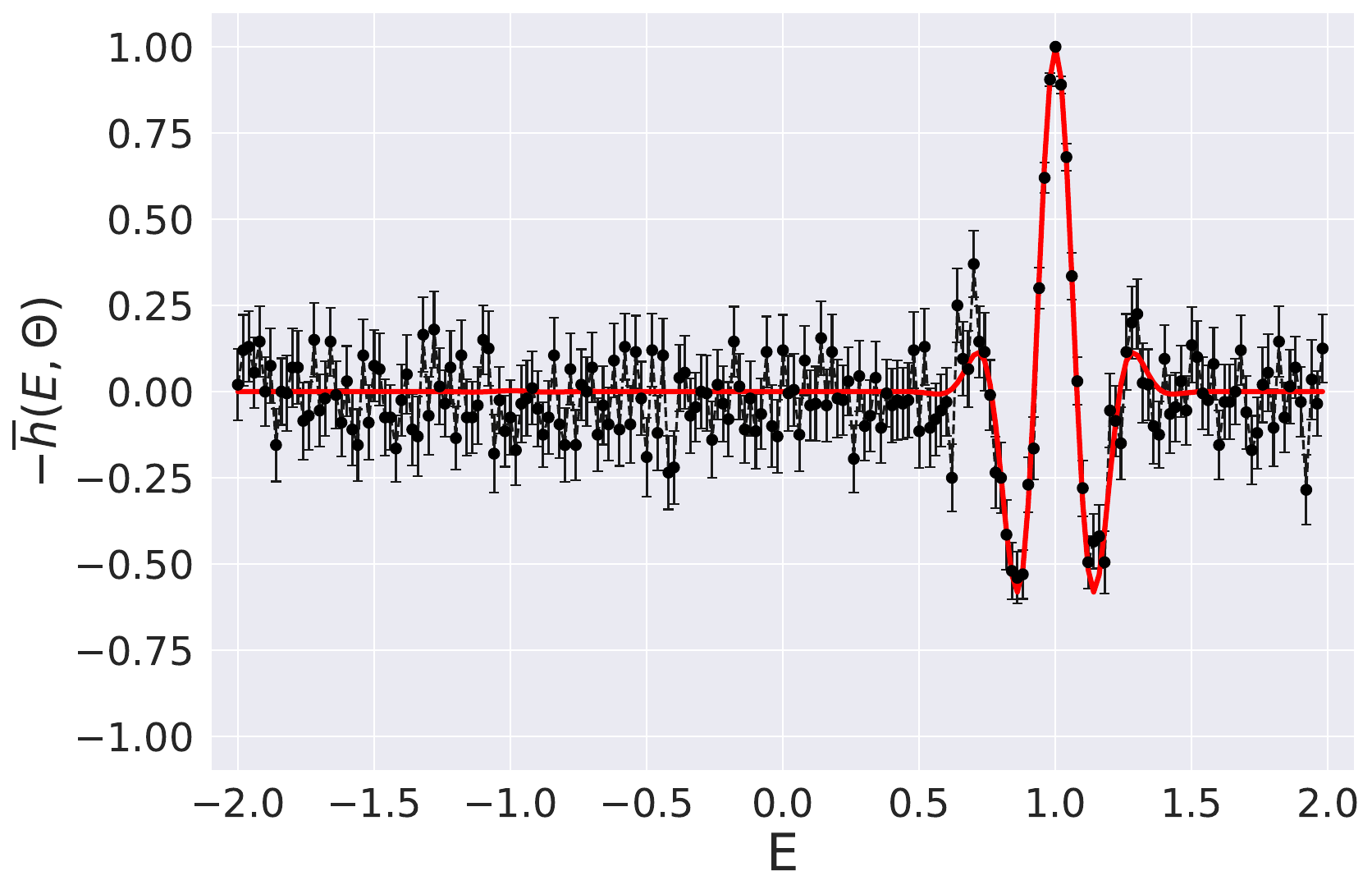}
    \includegraphics[scale=0.15]{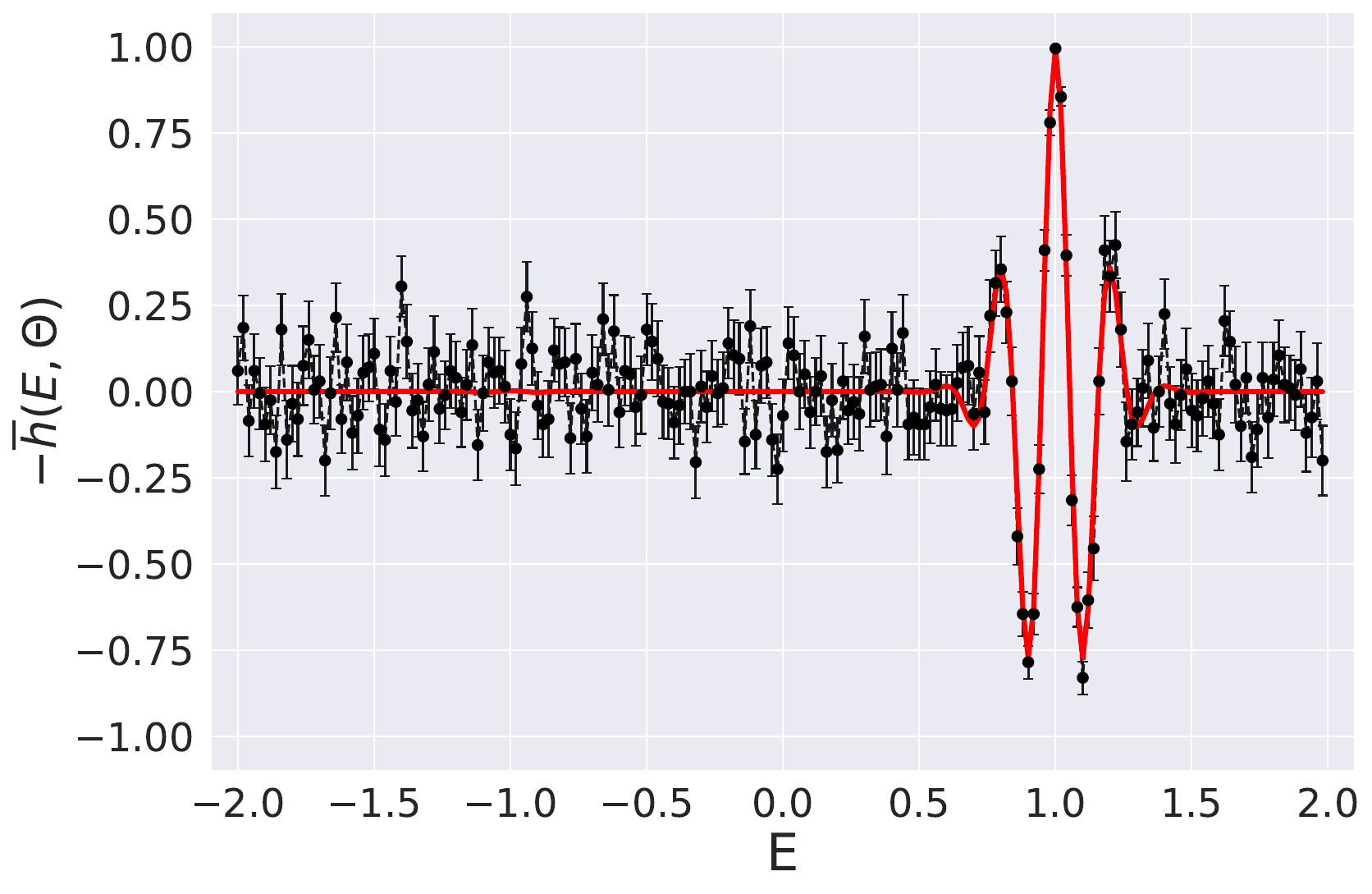}\\
    \caption{From top to down, specific results for the circuit build on Pennylane of the one-spin Zeeman model with $N=1$, $B=1.0$, $\theta=\pi/30$, $N_{\text{rounds}}=50$ and $d=7$ for $\tau=\{0;10;20;30\}$.}
\label{fig:plszn5e-020med10sigmod___NEW}
\end{figure}

The final strategies to refine the data set consist of tuning the remaining variables considered in Eq. (\ref{eq:ge}). In this sense, these changes can overcome the influence of the parameter $N$ and reduce the circuit's dependence on auxiliary qubits, expanding the possibilities to improve the results on NISQ devices.

The first choice is to find an appropriate standard deviation for the $t_{N}$ normal distribution through the parameters $d$ and $\tau$, which is crucial to understand the behavior of $-\overline {h}(E,\psi_I)$ near the Hamiltonian eigenvalues. Note in Fig. (\ref{fig:plszn5e-020med10sigmod}) that the peak becomes sharper around $E$ as $d$ increases, while the nearest points around $-\overline {h}(E,\psi_I)=1$ are set apart from it. On the other hand, despite finding the eigenvalues with more accuracy, increasing the value of $\tau$ demands a higher precision for setting the energy parameter on the $P(E t_{N})$ gate applied to each Rodeo cycle, which turns the measurement process more vulnerable to noise. In turn, Fig. \ref{fig:plszn5e-020med10sigmod___NEW} shows that the oscillations for the points at the peak basis are attenuated as $\tau$ approaches zero, due to the influence of the factor $\cos{\left[(E - E_x)\tau \right]}$ in Eq. (\ref{eq:ge}). For this reason, the best choices rely on set $d$ and $\tau \in [0;10$].
\begin{figure}[!ht]
   \includegraphics[scale=0.3]{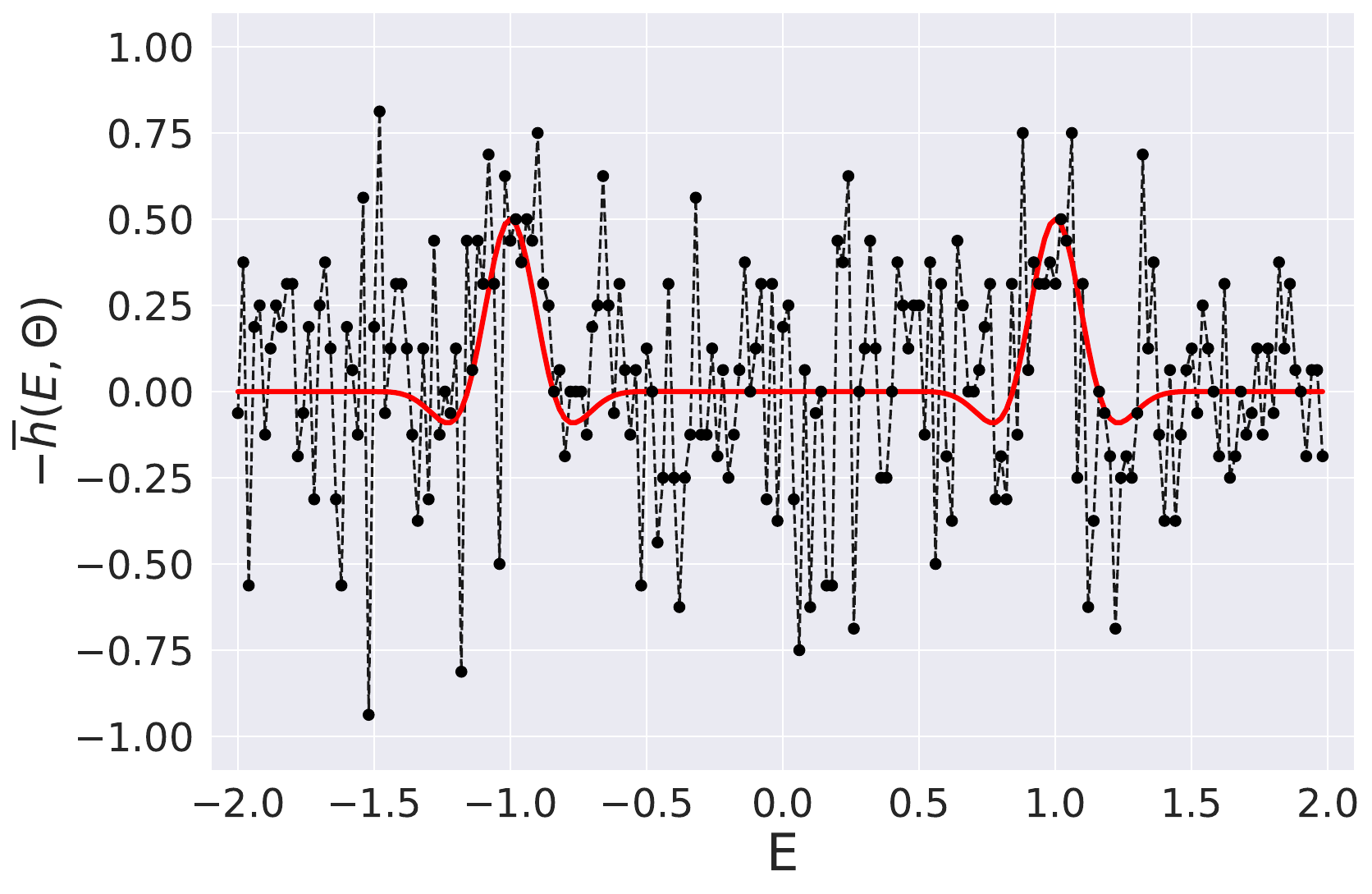}
    \includegraphics[scale=0.3]{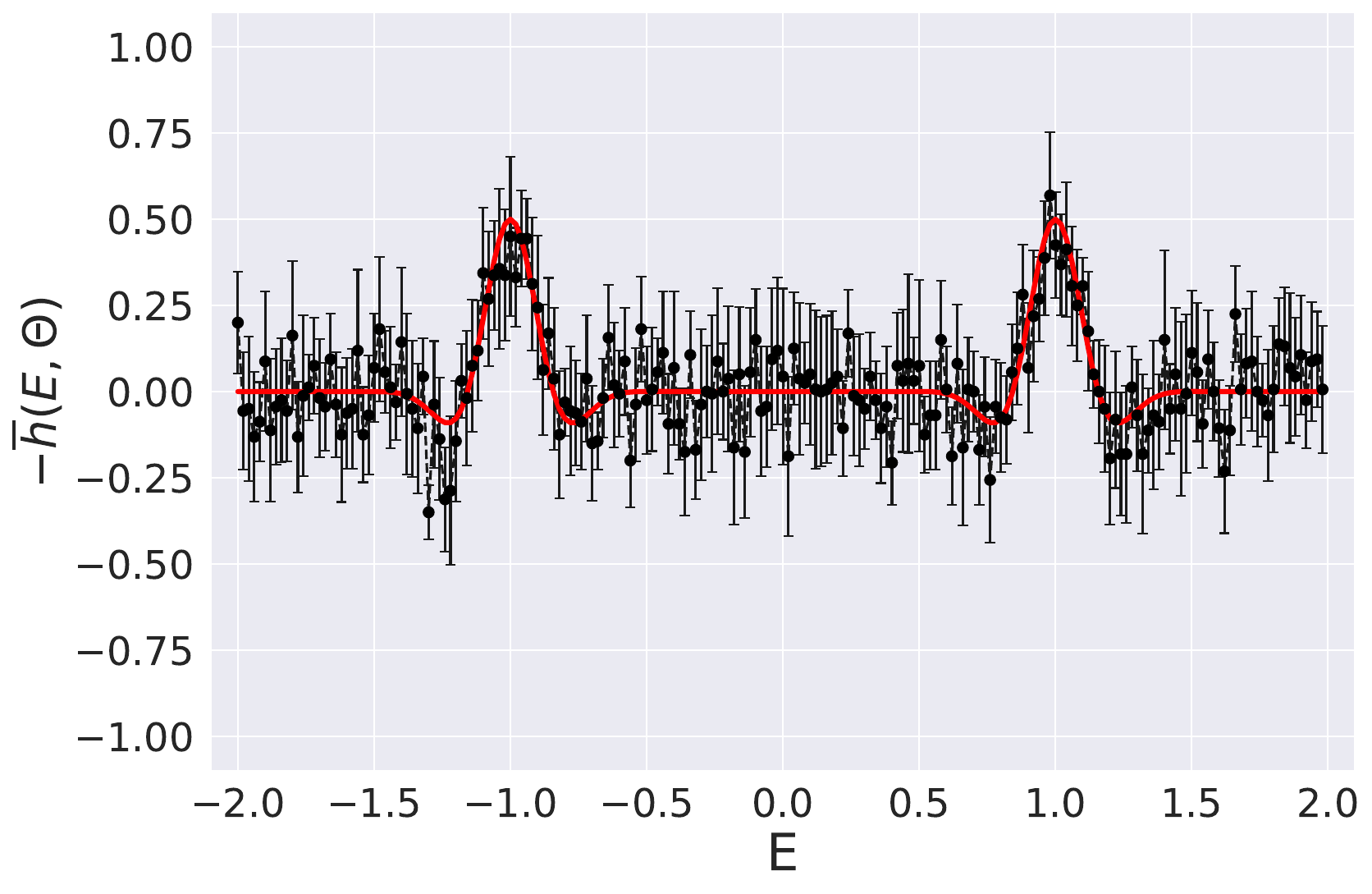}
    \includegraphics[scale=0.3]{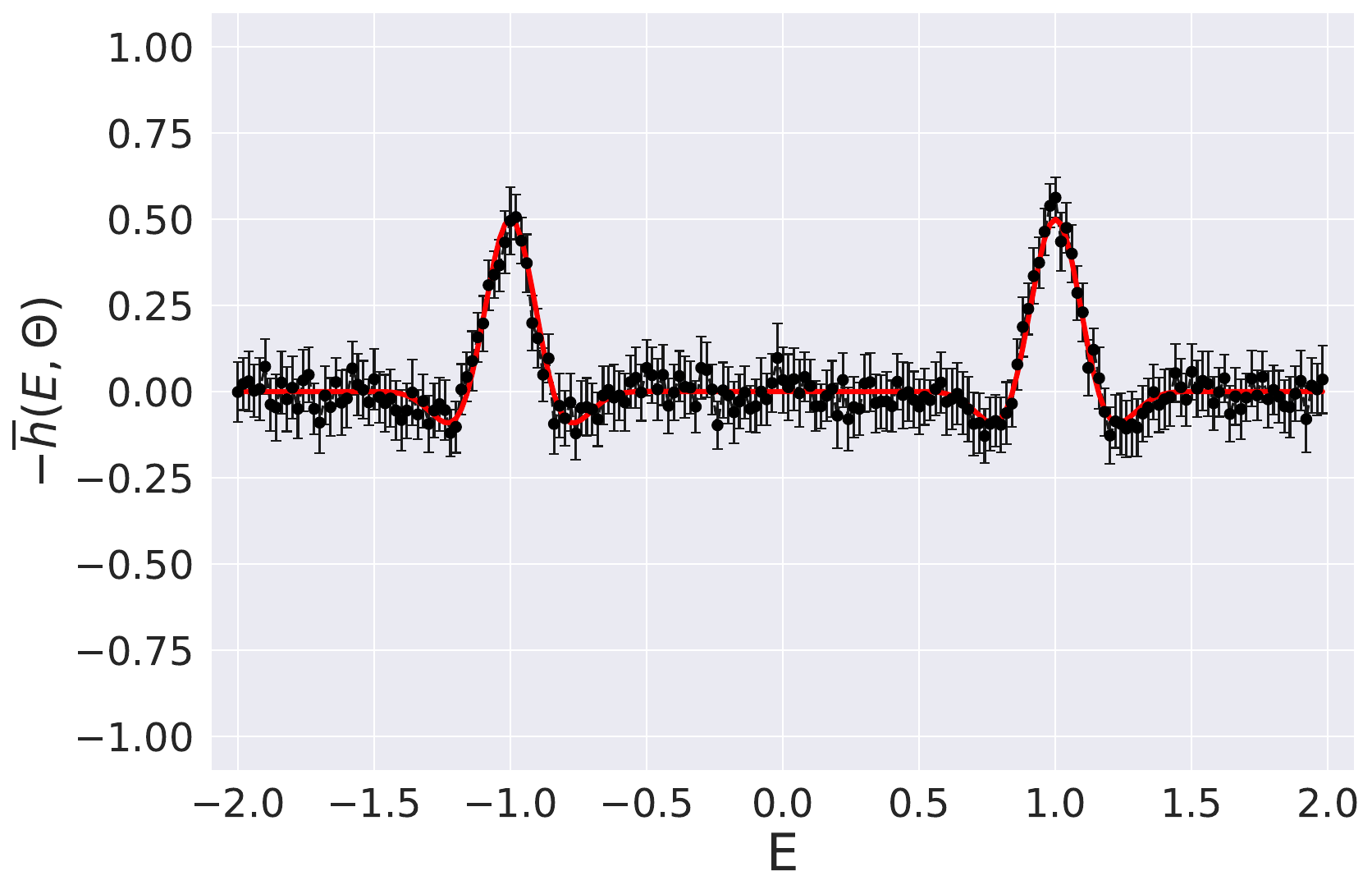} 
    \caption{From top to down, specific results for the circuit build on Pennylane of the one-spin Zeeman model with $N=1$, $B=1.0$, $\theta=\pi/2$, $N=4$, $d=7$ and $\tau=10$ for $N_{\text{rounds}}=\{1;10;50\}$.}
\label{fig:plszn5e-020rxmed1_10mod}
\end{figure}

The second choice is based on the definition of the initial state, which also plays a key role in the Rodeo Algorithm's success. When the overlap condition $-\overline {h}(E,\psi_I)\approx 1$ is not achieved, the results generated by Eq. (\ref{eq:getheta1}) may not be always sufficient to identify the eigenvalue spectrum of $H_{\text{obj}}$, and the strategies presented in this section become fundamental to determine the correct solutions. 

The first panel in Fig. \ref{fig:plszn5e-020rxmed1_10mod} shows the results obtained from a circuit simulated without measurement repetition ($N_{\text{rounds}}=1$) for $N=4$ and $\theta=\pi/2$.  
Although the points distribution around $-\overline {h}(E,\psi_I)\approx 0.5$ for both eigenvalues($E=\pm{1.00}$) approximates the slope, the uncertainty of this pattern cannot be ignored. On the other hand, the dispersion is refined in the center panel when the measurement repetition (with $N_{\text{rounds}}=10$) is included along the process, besides the advantage of reducing the circuit complexity. Finally, note that both eigenvalues become identifiable in the last panel for $N_{\text{rounds}}=50$.

If the properties of the initial state are completely unknown, all possible values of $\theta$ and $\phi$ must be tested on the rotations $R_x({\theta})$, $R_y({\theta})$ and $R_z({\phi})$ to maximize $-\overline {h}(E,\psi_I)$ for a specific $H_{\text{obj}}$. Consequently, the tuning processes allow us not only to create strategies to efficiently sample the entire rank of energy, but also pave the way to enhance the results through artificial intelligence techniques \cite{beer2020book,schuld2021book,glisic2022book,combarro2023book}. 
\subsection{ Two-Spin Hamiltonian}
\label{subsec:twospin}

The methodology presented so far can be extended to investigate whether the algorithm's predictions remain valid for multi-qubit systems with particular properties that have not yet been considered, such as degeneracy and entanglement. For this reason, we move forward to analyze the two-spin Zeeman model according to the coupling in the initial system. By choosing $M=2$ in Eq. (\ref{eq:ZemmanHam}), the Hamiltonian is now expressed by
\begin{eqnarray}
H_{\text{obj}} &=&-B (\sigma_{z}\otimes\mathds{1}+\mathds{1}\otimes\sigma_{z})\nonumber\\
        &=&-2B(\op{00}{00}-\op{11}{11}).
\label{eq:twospin}
\end{eqnarray}

From this point on, we will denote the computational basis for a bipartite system as $\ket{00}\equiv\ket{0}$, $\ket{01}\equiv\ket{1}$, $\ket{10}\equiv\ket{2}$ and $\ket{11}\equiv\ket{3}$. Note in Eq. (\ref{eq:twospin}) that the set of eigenvalues $E_0=-2B$, $E_1=E_2=0$ and $E_3=2B$ is directly associated to the standard basis, considering that the states $\ket{1}$ and $\ket{2}$ are degenerated. 

In the following topics, we apply the Rodeo Algorithm to explore two scenarios whose dynamics are governed by Eq. (\ref{eq:twospin}): one in which $\ket{\psi_I}$ represents a pair of non-interacting spins and other where the circuit starts with an entangled state.
\subsubsection{Non-interacting states}
\label{subsubsec:separablestates}

The target system defined in Eq. (\ref{eq:SN1G}) can be expanded as a tensorial product of two arbitrary qubits that do not interact with each other, which leads to
\begin{eqnarray}
\ket{\psi_I}&=& \cos{\left(\frac{\theta_1}{2}\right)}\cos{\left(\frac{\theta_2}{2}\right)}\ket{00}   \nonumber\\
&+& e^{i\varphi_1}\sin{\left(\frac{\theta_1}{2}\right)}\cos{\left(\frac{\theta_2}{2}\right)}\ket{10}\nonumber\\
&+&e^{i\varphi_2}\cos{\left(\frac{\theta_1}{2}\right)}\sin{\left(\frac{\theta_2}{2}\right)}\ket{01}\nonumber\\ 
&+& e^{i(\varphi_1+\varphi_2)}\sin{\left(\frac{\theta_1}{2}\right)}\sin{\left(\frac{\theta_2}{2}\right)}\ket{11}, 
\label{eq:SN2G}
\end{eqnarray}
where $\theta_i$ and $\varphi_i$ for $i \in \{1,2\}$ are respectively the polar and azimuthal angles in the Bloch sphere of the $i$-th qubit.
\begin{figure}[!ht]
\centering	\includegraphics[scale=0.21]{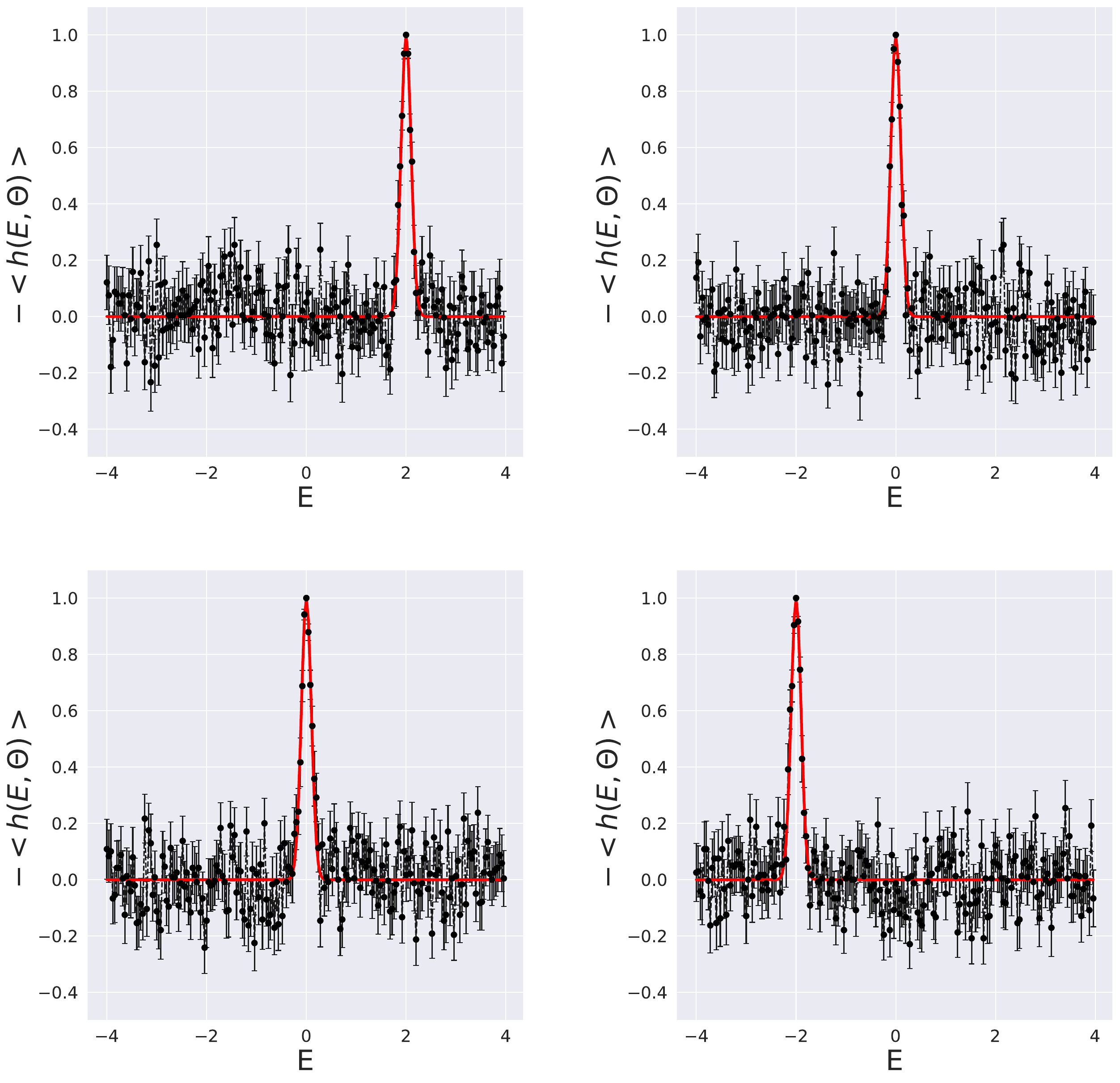}
        \caption{ From top to down, specific results for the circuit build on Pennylane of the two-spin Zeeman model with $N=1$, $B_z=1.0$, $\tau=0$, $d=10$ and $N_{\text{rounds}}=60$ for $\theta_{1,2}=\{0;0\};\;\{0;\pi\};
     ;\{\pi;0\};\;\{\pi;\pi\}$.  }
\label{fig:gEZ2}
\end{figure}

Following the conventions adopted in subsection
\ref{subsec:onespin} for the one-spin Zeeman model, we set $\varphi_1 = \varphi_2 = 0$ to express the Hamiltonian only in terms of $\theta_1$ and $\theta_2$. Thereby, Eq. (\ref{eq:ge}) results in
\begin{eqnarray}
&&\overline{h}(E,\theta_1,\theta_2) = \nonumber\\
&-& \cos^{2}{\left( \frac{\theta_1}{2} \right)} \cos^{2}{\left( \frac{\theta_2}{2} \right)} e^{-\frac{d^2 (E+2B)^2}{2}}\cos{[(E+2B)\tau]} \nonumber\\
&-&   \cos^{2}{\left( \frac{\theta_1}{2} \right)} \sin^{2}{\left( \frac{\theta_2}{2} \right)}e^{-\frac{d^2  E^2}{2}}\cos{(E\tau)} \nonumber\\
&-& \sin^{2}{\left( \frac{\theta_1}{2} \right)} \cos^{2}{\left( \frac{\theta_2}{2} \right)}e^{-\frac{d^2  E^2}{2}}\cos{(E\tau)}  \\
&-& \sin^{2}{\left( \frac{\theta_1}{2} \right)} \sin^{2}{\left( \frac{\theta_2}{2} \right)}e^{-\frac{d^2 (E-2B)^2}{2}}\cos{[(E-2B)\tau]},\nonumber
\label{eq:ge1medt1t2}
\end{eqnarray}
where the set of rotations $\{\theta_1;\theta_2\}$ associated to the computational basis provides the maximum value $-\overline{h}(E_{j},\theta_1,\theta_2)=1$ for $j=\{0,1,2,3\}$.

Fig. \ref{fig:gEZ2} shows the results for $N=1$, $B=1.0$, $d=10$, $\tau=0$ and $N_{\text{rounds}}=60$. Note that all peaks are well-defined around the predicted eigenvalues, with the degeneracy regarding $E_{1}=E_{2}=0$ appearing twice. Since there is no spin correlation in Eqs. (\ref{eq:twospin}) and (\ref{eq:SN2G}), the method effectively fulfills the task of revealing the eigenvalues
and directing $\ket{\psi_{I}}$ to the respectively eigenstates $\ket{0}$, $\ket{1}$, $\ket{2}$ and $\ket{3}$. 
\subsubsection{Entangled states
\label{subsubsec:entangledstates}}

We now assume that the initial state is maximally entangled. Since the Bell basis can be expressed in terms of the computational one \cite{bell1964} through the following relation:
\begin{eqnarray}
\ket{\Phi^{+}} = \frac{1}{\sqrt{2}}\left(\ket{00}+\ket{11} \right);\;
\ket{\Phi^{-}} = \frac{1}{\sqrt{2}}\left(\ket{00}- \ket{11}\right);\nonumber\\
\ket{\Psi^{+}} = \frac{1}{\sqrt{2}}\left(\ket{01}+\ket{10} \right);\;
\ket{\Psi^{-}} = \frac{1}{\sqrt{2}}\left(\ket{01}-\ket{10} \right);\nonumber\\
\label{eq:bellbasis}
\end{eqnarray}
it is straightforward to define $\ket{\psi_I}$ as an entangled state from Eq. (\ref{eq:bellbasis}). 

Let us consider inputs initially described by $\ket{\Phi^{+}}$ or $\ket{\Phi^{-}}$. The probability outcomes for $\ket{00}$ or $\ket{11}$ after a measurement on the computational basis for both cases are equally likely, once the von Neumann entropy is maximum \cite{Neumann1955book}.
In fact, Eq.(\ref{eq:ge}) results in
\begin{eqnarray}
\overline{h}(E,\Phi^{+}) &=& \overline{h}(E,\Phi^{-})=\nonumber\\
 &-& \frac{1}{2}e^{-\frac{d^2 (E+2B)^2}{2}}\cos{[(E+2B)\tau]} \nonumber\\
&-&\frac{1}{2}e^{-\frac{d^2N (E-2B)^2}{2}}\cos{[(E-2B)\tau]},
\label{eq:phimaismenos}
\end{eqnarray}
which has two global maxima for $-\overline{h}(E=\pm 2B,\Phi^{\pm})=0.5$, as shown in the left panel of Fig. \ref{fig:gE_bell}. Since each wave function describes a superposition of the eigenstates $\ket{0}$ and $\ket{3}$, both solutions can be obtained through the methods discussed in subsection \ref{subsec:onespin}.

One would expect to mirror this choice to find the remaining solutions by setting the initial state as $\ket{\Psi^{+}}$ or $\ket{\Psi^{-}}$. However, as the eigenstates $\ket{1}$ and $\ket{2}$ are degenerated, their distributions can not be distinguishable from each other. For $\ket{\psi_I}=\ket{\Psi^{\pm}}$, Eq.(\ref{eq:ge}) leads to
\begin{equation}
 \overline{h}(E,\Psi^{+}) = \overline{h}(E,\Psi^{-})= 
 -e^{-\frac{d^2 E^2}{2}}\cos{(E\tau)},
\label{eq:psimaismenos}
\end{equation}
whose global maxima is achieved exactly for $-\overline{h}(E=0,\Psi^{\pm})=1$. 
\begin{figure}[!ht]
	\centering	\includegraphics[scale=0.2]{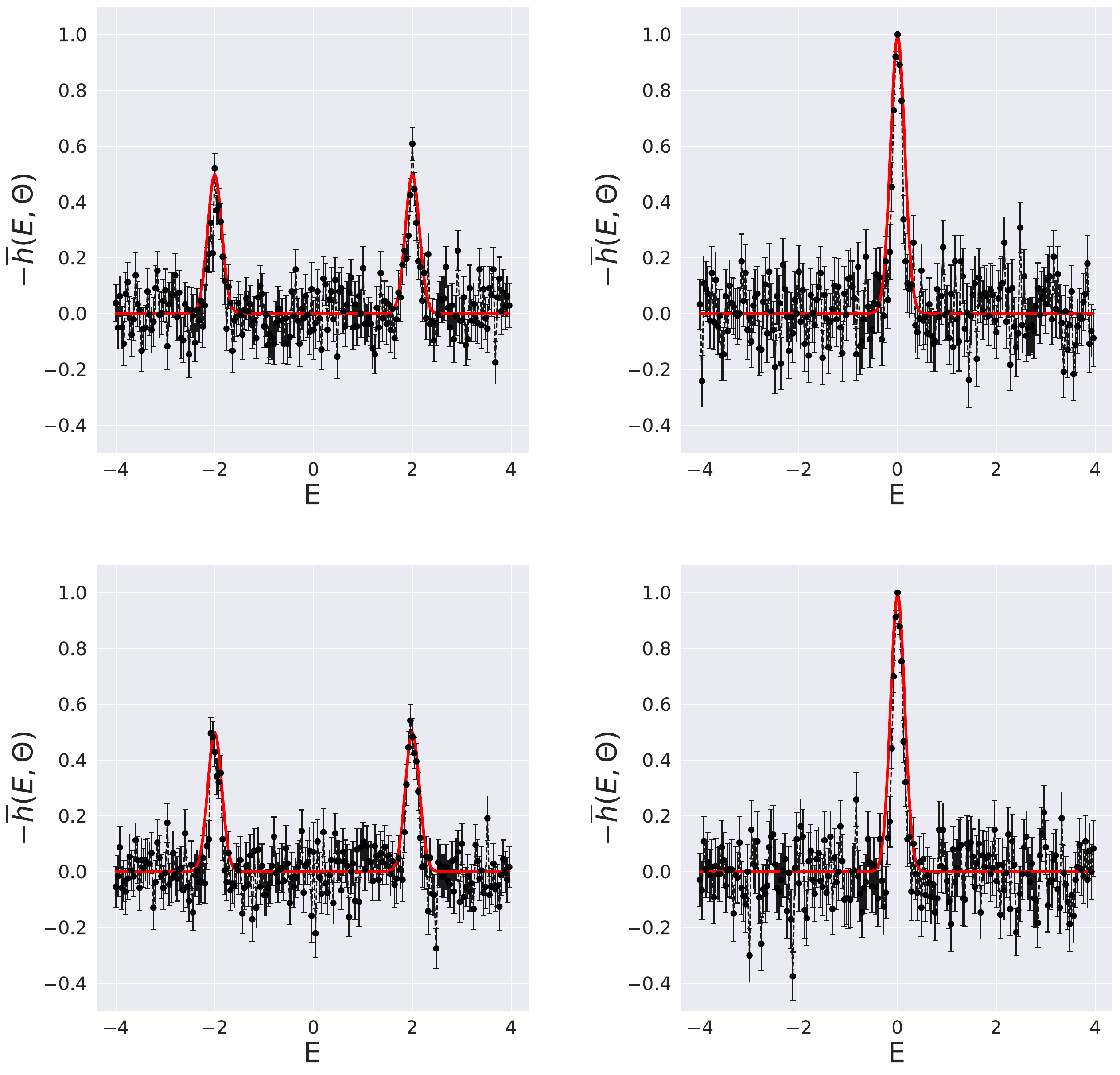} 
        \caption{  From top to down, specific results for the circuit build on Pennylane of the two-spin Zeeman model with $N=1$, $B=1.0$, $\tau=0$, $d=10$ and $N_{\text{rounds}}=60$ for $\ket{\psi_I}=\{\ket{\Phi^{+}};\ket{\Psi^{+}};\ket{\Phi^{-}};\ket{\Psi^{-}}\}$. }
\label{fig:gE_bell}
\end{figure}

The right panels in Fig. \ref{fig:gE_bell} corroborate this hypothesis: in both cases, the slopes merge in a single peak centered on $E=0$ that reaches the maximum value at once. Moreover, since any linear distribution of degenerate eigenstates is also an eigenstate, the set of all possible combinations of $\ket{\Psi^{+}}$ and $\ket{\Psi^{-}}$ (including the inputs $\ket{01}$ and $\ket{10}$) gives the same result. 

Likewise, the same arguments remain valid when the initial state or the eigenstates are not maximally entangled. For instance, if $\ket{\psi_I}$ is defined as a partially entangled state \cite{fortes2015} for an arbitrary real parameter $\alpha$ as
\begin{equation}
\ket{\psi_I}=\cos(\alpha)\ket{\Psi^{+}}+\sin(\alpha)\ket{\Psi^{-}},
\label{eq:lincomb_psi}
\end{equation}
Eq.(\ref{eq:ge}) results in
\begin{equation}
\overline{h}(E,\psi_I) =
 -e^{-\frac{d^2 E^2}{2}}\cos{(E\tau)},
\end{equation}
whose prediction is exactly the same as the one given by Eq. (\ref{eq:psimaismenos}). On the other hand, for
\begin{equation}
\ket{\psi_I}=\cos(\alpha)\ket{\Phi^{+}}+\sin(\alpha)\ket{\Phi^{-}},
\label{eq:lincomb_phi}
\end{equation}
Eq.(\ref{eq:ge}) is expressed by
\begin{eqnarray}
\overline{h}(E,\psi_I) &=&  \nonumber\\
 &-&\cos(\alpha)^2 e^{-\frac{d^2 (E+2B)^2}{2}}\cos{[(E+2B)\tau]} \nonumber\\
&-&\sin(\alpha)^2e^{-\frac{d^2N (E-2B)^2}{2}}\cos{[(E-2B)\tau]},
\label{eq:lincomb_phimaismenos}
\end{eqnarray}
which generates two peaks with distinct amplitudes for $\alpha \neq \pi/4$.
\begin{figure}[!ht]
	\centering	\includegraphics[scale=0.2]{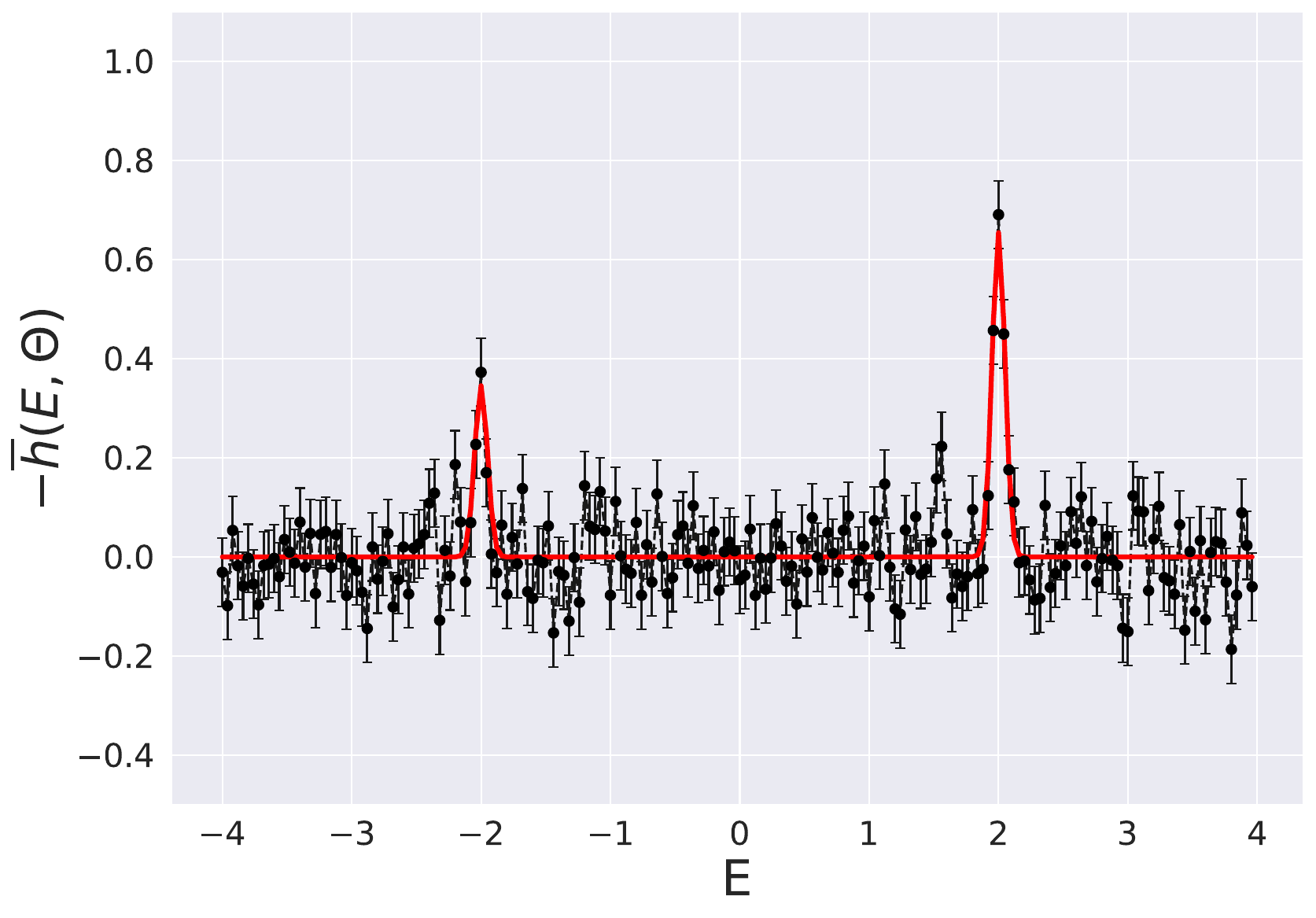} 
        \caption{From top to down, specific results for the circuit build on Pennylane of the two-spin Zeeman model with $N=1$, $B=1.0$, $\tau=0$, $d=20$ and $N_{\text{rounds}}=60$ for $\ket{\psi_I}=\cos(\pi/5)\ket{\Phi^{+}}+\sin(\pi/5)\ket{\Phi^{-}}$.}
\label{fig:gE_lin_comb_bell}
\end{figure}

Fig. \ref{fig:gE_lin_comb_bell} presents the results obtained for $\alpha=\pi/5$. As expected, the peaks satisfies the condition $P(00)+ P(11)\approx1$, where $P(11) \approx -\overline{h}(E=+2.0,\theta= \pi/5) \approx 0.66$ and $P(00) \approx -\overline{h}(E=-2.0,\theta=\pi/5) \approx 0.35$, analogously to pattern exhibited in the last panel of Fig. \ref{fig:gE1}. Therefore, the strategies proposed here can also be employed in future studies to characterize an arbitrary Hamiltonian in circuits with multiple qubits associated with pure states.

A similar procedure can be adopted if the spins are coupled due to the Hamiltonian interaction in Eq. (\ref{eq:twospin}), considering that the eigenstates are entangled. In this case, the previous analysis must be inverted, so that the initial system can be defined in terms of the computational basis (according to the respective eigenstates' degeneracy).

In the next section, we present the last results by implementing the Rodeo Algorithm on a real superconducting device.
\section{Results obtained on a real device}
\label{sec:results-dev}
\begin{figure}[!ht]
    \includegraphics[scale=0.25]{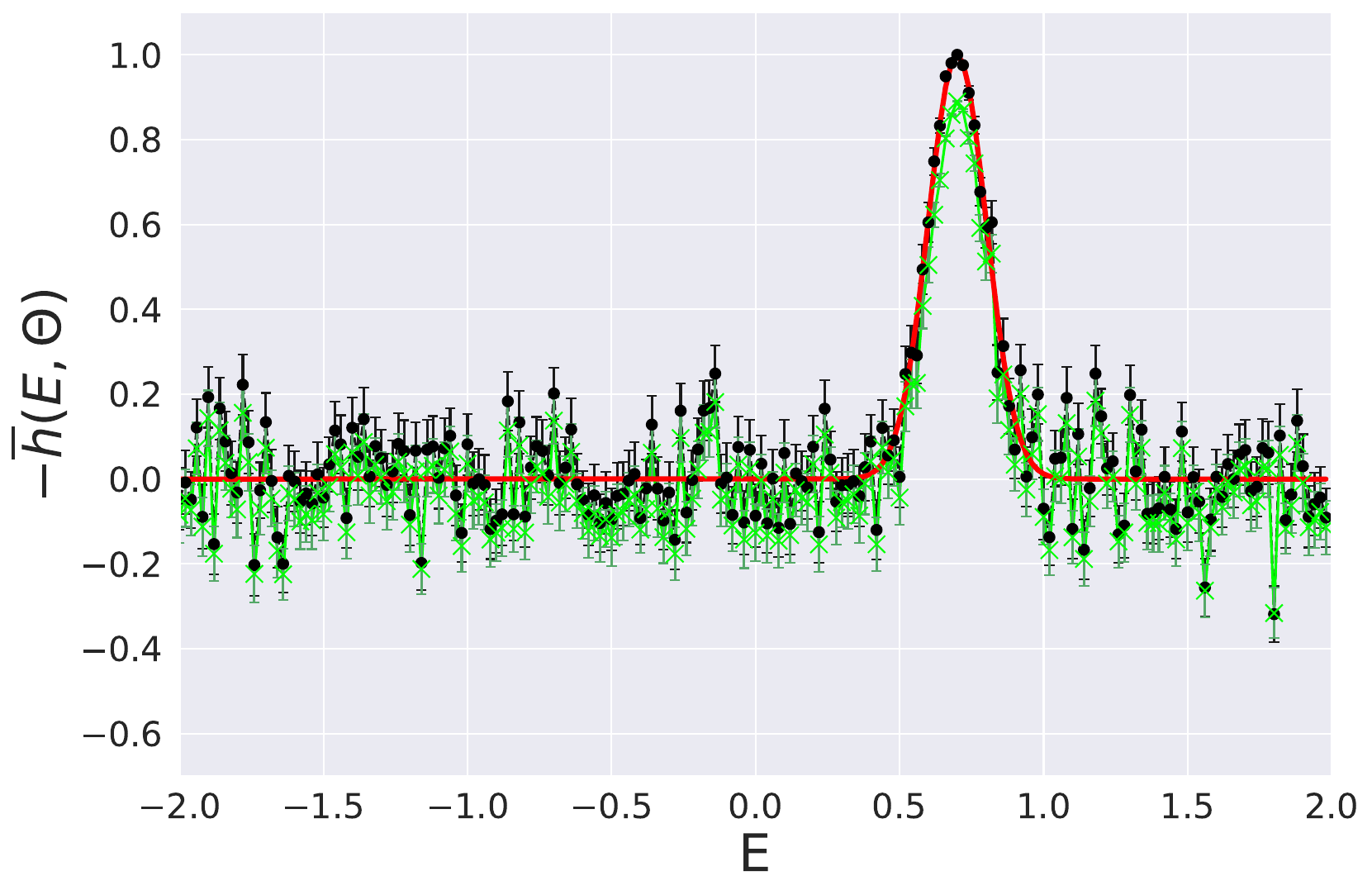}\\
    \includegraphics[scale=0.25]{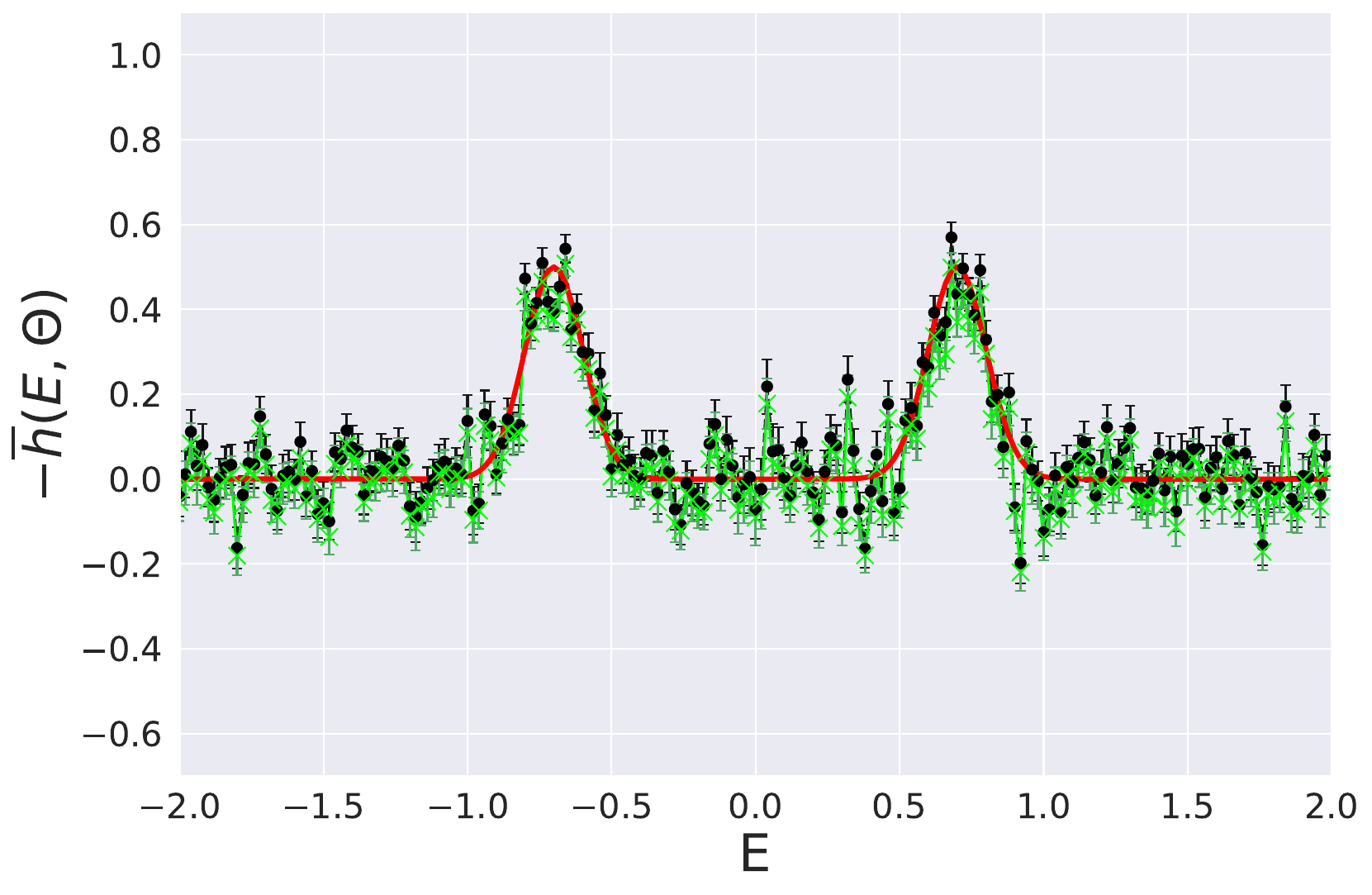}\\
    \includegraphics[scale=0.25]{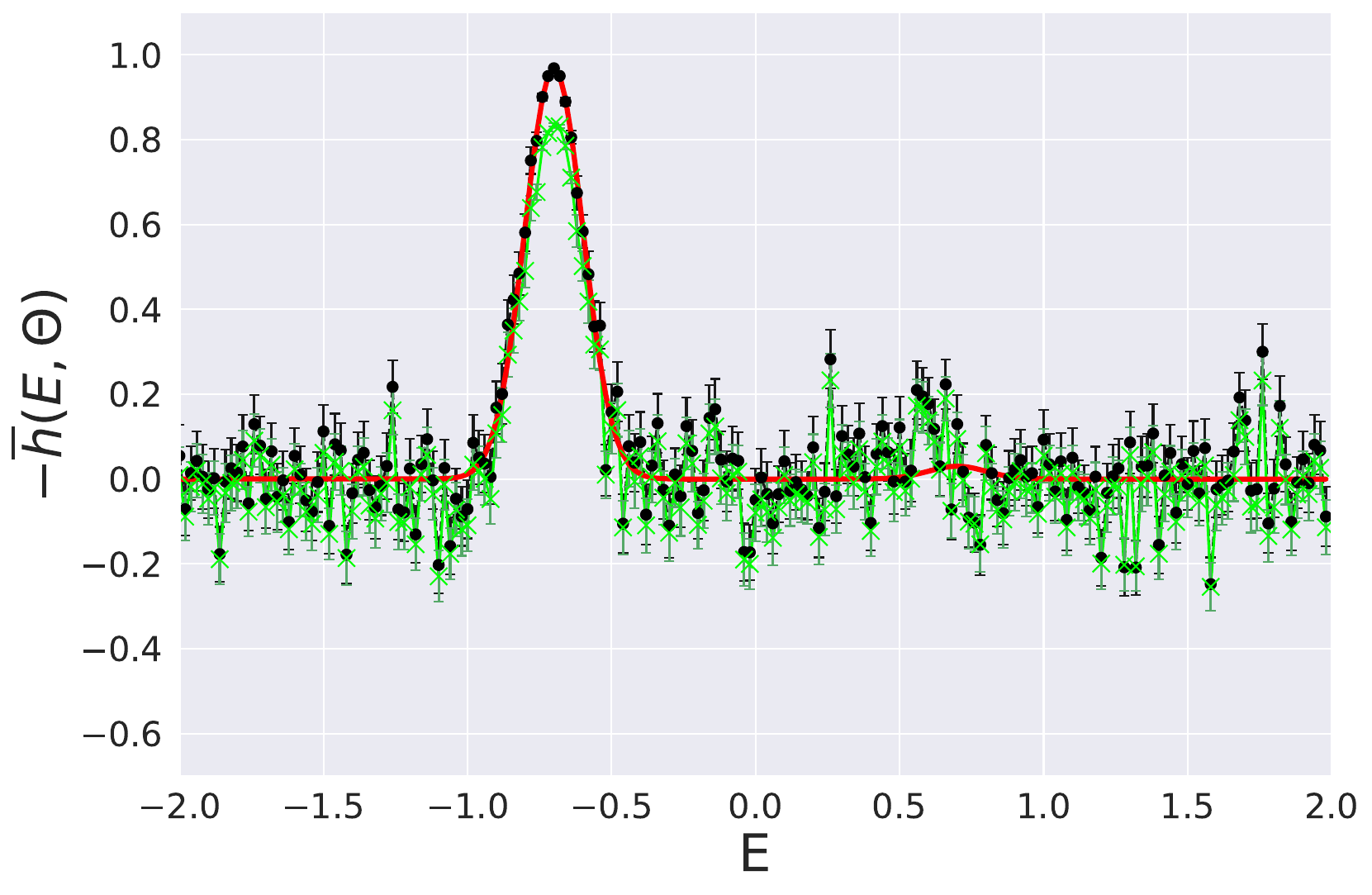}
    \caption{ From top to down, specific results obtained through the $ibmq\_lima$ device (from IBM Qiskit) of the one-spin Zeeman model with $N=1$, $B=0.7$, $\tau=0$, $d=7$ and $N_{\text{rounds}}=50$ for $\theta=\{0;\pi/2;\pi\}$.
\label{fig:gEQ}}
\end{figure}
The Qiskit platform is one of the most widespread open-source software development kits currently available for quantum computing. Unlike Pennylane, this interface allows the user to directly transpose the codes from simulators onto real devices. For this reason, we also implement the Rodeo Algorithm on it
to show that the method is platform-independent. 

We choose the spin Zeeman model presented in Eq. (\ref{eq:onespin}) for comparison purposes. According to the data set introduced in Eq.(\ref{eq:X_iY_i}), the expected value for each qubit is evaluated as  
\begin{equation}
\langle\sigma_z\rangle_{k}=\frac{(N_{\uparrow}-N_{\downarrow})}{(N_{\uparrow}+N_{\downarrow})},
\label{eq:expvalibm}
\end{equation}
where $N_{\uparrow}$ and $N_{\downarrow}$ represent the respective outcomes corresponding to measurements on the computational basis. 

Fig. \ref{fig:gEQ} reproduce the results acquired by the $ibmq\_lima$ device \textcolor{blue}{\cite{refqiskit}}. Note that the new distribution (highlighted in green) exhibits a similar pattern as the one generated in Pennylane, except for a particular aspect: due to the decoherence process inherent in any real device, the function never reaches the point $-\overline {h}(E_{x},\psi_I)=1$. 
\begin{figure}[!ht]
    \includegraphics[scale=0.45]{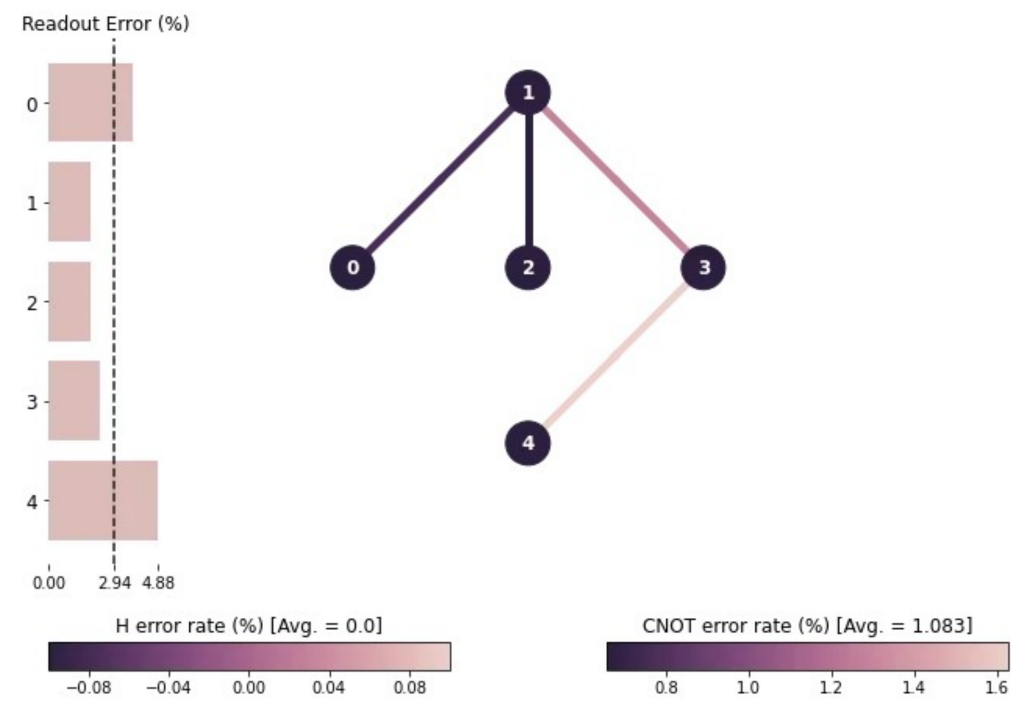}
    \caption{$ibmq_lima$ quantum processor error rates for H and CNOT gates \cite{mapibmqlima}
\label{fig:ibmqlima}}
\end{figure}

It is worth noting that this noise comes mainly from the quantum gates error rates and the connectivity between the qubits, as shown in Fig. \ref{fig:ibmqlima}. Since the map is not fully connected, the device allocates extra $SWAP$ gates to perform the information processing required by the theoretical circuit, which also increases its space and time complexity. However, even with this unavoidable limitation, the peaks are slightly reduced by the loss for the proposed model, and the results can be improved by the analysis and strategies presented in section \ref{sec:results-sim}. 
\section{Conclusions}
\label{sec:conclusion}

We managed to unravel the Rodeo Algorithm by applying it to single and bipartite systems described by the Zeeman model through the Pennylane and Qiskit platforms. By introducing the Rider state and the Bull operator to develop a new methodology based on numerical and graph analysis, we not only detail the procedure for characterizing the spectrum of eigenstates and eigenvalues of an arbitrary Hamiltonian (showing that the eigenvalues can be inferred simultaneously) but also improve the method by eliminating the need to have prior knowledge about these properties and the initial state $|\psi_{I}\rangle$. In this context, we presented strategies and techniques to refine the results by tuning the parameters to reduce the fluctuations inherent to data distribution, as well as to improve the algorithm's performance.  

We also explored scenarios where the presence of entanglement and degeneracy creates new possibilities to understand the pattern for bipartite systems, indicating that the strategies proposed here can be extended to multiple qubits. The results reveal that the algorithm performance also depends on the characteristics of the Hamiltonian. If the spins do not interact and the eigenvalues are unique, the choice to start with an entangled state is more advantageous (and vice-versa), due to the superposition between the solutions sought. Hence, in order to define a set of eigenstates compatible with the respective model, it is necessary not only to sample the entire configuration space for $\ket{\psi_I}$ (including its entanglement degree), but also to compare the results with the ones obtained for the non-degenerate cases to understand the system dynamics.

Finally, we compare the predictions generated by simulators for a single qubit with the results obtained on the $ibmq\_lima$ device, confirming the theoretical predictions even in the presence of noise. Most importantly, we show the proper operation of the technique on a real quantum computer. For future work, we intend to build larger data sets and test Eq. (\ref{eq:expvalibm}) to explore scenarios for multi-qubit systems.
\newpage

\bibliographystyle{ieeetr}  
\bibliography{references}  

\begin{thebibliography}{10}

\bibitem{cohen2019book}
C.~Cohen-Tannoudji, B.~Diu, and F.~Lalo{\"e}, {\em Quantum Mechanics, Volume 1:
  Basic Concepts, Tools, and Applications}.
\newblock Wiley, 2019.

\bibitem{sakurai2017book}
J.~Sakurai and J.~Napolitano, {\em Modern Quantum Mechanics}.
\newblock Cambridge University Press, 2017.

\bibitem{nielsen2010book}
M.~Nielsen and I.~Chuang, {\em Quantum Computation and Quantum Information:
  10th Anniversary Edition}.
\newblock Cambridge University Press, 2010.

\bibitem{rigol2008}
M.~Rigol, V.~Dunjko, and M.~Olshanii, ``Thermalization and its mechanism for
  generic isolated quantum systems,'' {\em Nature}, vol.~452, pp.~854--8, Apr
  2008.

\bibitem{dirac1981book}
P.~A.~M. Dirac, {\em The principles of quantum mechanics}.
\newblock Oxford university press, 1981.

\bibitem{modisette1996}
J.~Modisette, P.~Nordlander, J.~Kinsey, and B.~Johnson, ``Wavelet based in
  eigenvalue problems in quantum mechanics,'' {\em Chemical Physics Letters},
  vol.~250, pp.~485--494, Mar 1996.

\bibitem{polizzi2009}
E.~Polizzi, ``Density-matrix-based algorithm for solving eigenvalue problems,''
  {\em Phys. Rev. B}, vol.~79, p.~115112, Mar 2009.

\bibitem{beugeling2018}
W.~Beugeling, A.~B\"acker, R.~Moessner, and M.~Haque, ``Statistical properties
  of eigenstate amplitudes in complex quantum systems,'' {\em Phys. Rev. E},
  vol.~98, p.~022204, Aug 2018.

\bibitem{schutt2019}
K.~Schütt, M.~Gastegger, A.~Tkatchenko, K.-R. Müller, and R.~Maurer,
  ``Unifying machine learning and quantum chemistry with a deep neural network
  for molecular wavefunctions,'' {\em Nature Communications}, vol.~10, p.~5024,
  Nov 2019.

\bibitem{landau2021book}
D.~Landau and K.~Binder, {\em A Guide to Monte Carlo Simulations in Statistical
  Physics}.
\newblock Cambridge University Press, 2021.

\bibitem{feynman1982}
R.~P. Feynman, ``Simulating physics with computers,'' {\em International
  Journal of Theoretical Physics}, vol.~21, pp.~467--488, June 1982.

\bibitem{lloyd1996}
S.~Lloyd, ``Universal quantum simulators,'' {\em Science}, vol.~273,
  pp.~1073--1078, Aug 1996.

\bibitem{lloyd1998}
S.~Lloyd, ``Universal quantum simulators: Correction,'' {\em Science},
  vol.~279, pp.~1113--1117, Feb 1998.

\bibitem{abrams1999}
D.~S. Abrams and S.~Lloyd, ``Quantum algorithm providing exponential speed
  increase for finding eigenvalues and eigenvectors,'' {\em Phys. Rev. Lett.},
  vol.~83, pp.~5162--5165, Dec 1999.

\bibitem{buluta2009}
I.~Buluta and F.~Nori, ``Quantum simulators,'' {\em Science}, vol.~326,
  pp.~108--111, Oct 2009.

\bibitem{brown2010}
K.~Brown, W.~Munro, and V.~Kendon, ``Using quantum computers for quantum
  simulation,'' {\em Entropy}, vol.~12, pp.~2268--2307, Nov 2010.

\bibitem{aaronson2011}
S.~Aaronson and A.~Arkhipov, ``The computational complexity of linear optics,''
  in {\em Proceedings of the forty-third annual ACM symposium on Theory of
  computing}, pp.~333--342, June 2011.

\bibitem{georgescu2014}
I.~M. Georgescu, S.~Ashhab, and F.~Nori, ``Quantum simulation,'' {\em Rev. Mod.
  Phys.}, vol.~86, pp.~153--185, Mar 2014.

\bibitem{cerezo2021}
M.~Cerezo, A.~Arrasmith, R.~Babbush, S.~C. Benjamin, S.~Endo, K.~Fujii, J.~R.
  McClean, K.~Mitarai, X.~Yuan, L.~Cincio, {\em et~al.}, ``Variational quantum
  algorithms,'' {\em Nature Reviews Physics}, vol.~3, pp.~625--644, Aug 2021.

\bibitem{preskill2018}
J.~Preskill, ``Quantum computing in the nisq era and beyond,'' {\em Quantum},
  vol.~2, p.~79, Aug 2018.

\bibitem{peruzzo2014}
A.~Peruzzo, J.~McClean, P.~Shadbolt, M.-H. Yung, X.-Q. Zhou, P.~J. Love,
  A.~Aspuru-Guzik, and J.~L. O’brien, ``A variational eigenvalue solver on a
  photonic quantum processor,'' {\em Nature communications}, vol.~5, p.~4213,
  July 2014.

\bibitem{farhi2014}
E.~Farhi, J.~Goldstone, and S.~Gutmann, ``A quantum approximate optimization
  algorithm,'' {\em arXiv:1411.4028v1 [quant-ph]}, Nov 2014.

\bibitem{choi2021}
K.~Choi, D.~Lee, J.~Bonitati, Z.~Qian, and J.~Watkins, ``Rodeo algorithm for
  quantum computing,'' {\em Phys. Rev. Lett.}, vol.~127, p.~040505, July 2021.

\bibitem{sherbert2022}
K.~M. Sherbert, N.~Naimipour, H.~Safavi, H.~C. Shaw, and M.~Soltanalian,
  ``Quantum compressive sensing: Mathematical machinery, quantum algorithms,
  and quantum circuitry,'' {\em Applied Sciences}, vol.~12, p.~7525, July 2022.

\bibitem{guzman2022}
E.~A.~R. Guzman and D.~Lacroix, ``Accessing ground-state and excited-state
  energies in a many-body system after symmetry restoration using quantum
  computers,'' {\em Phys. Rev. C.}, vol.~105, p.~024324, Feb 2022.

\bibitem{pederiva2021}
G.~Pederiva, A.~Bazavov, B.~Henke, L.~Hostetler, D.~Lee, H.-W. Lin, and
  A.~Shindler, ``Quantum state preparation for the schwinger model,'' {\em
  arXiv:2109.11859v1 [hep-lat]}, Sep 2021.

\bibitem{perez2022}
E.~A. Coello~P\'erez, J.~Bonitati, D.~Lee, S.~Quaglioni, and K.~A. Wendt,
  ``Quantum state preparation by adiabatic evolution with custom gates,'' {\em
  Phys. Rev. A}, vol.~105, p.~032403, Mar 2022.

\bibitem{stetcu2022}
I.~Stetcu, A.~Baroni, and J.~Carlson, ``Variational approaches to constructing
  the many-body nuclear ground state for quantum computing,'' {\em Phys. Rev.
  C}, vol.~105, p.~064308, Jun 2022.

\bibitem{qian2024}
Z.~Qian, J.~Watkins, G.~Given, J.~Bonitati, K.~Choi, and D.~Lee,
  ``Demonstration of the rodeo algorithm on a quantum computer,'' {\em
  arXiv:2110.07747v2 [quant-ph]}, July 2024.

\bibitem{cohen2023}
T.~D. Cohen and H.~Oh, ``Optimizing the rodeo projection algorithm,'' {\em
  Phys. Rev. A}, vol.~108, p.~032422, Sep 2023.

\bibitem{lindgren2024}
M.~Bee-Lindgren, Z.~Qian, M.~DeCross, N.~C. Brown, C.~N. Gilbreth, J.~Watkins,
  X.~Zhang, and D.~Lee, ``Controlled gate networks applied to eigenvalue
  estimation,'' {\em arXiv:2208.13557v3 [quant-ph]}, 2024.

\bibitem{refpennylane}
V.~Bergholm, J.~Izaac, M.~Schuld, C.~Gogolin, S.~Ahmed, V.~Ajith, M.~S. Alam,
  G.~Alonso-Linaje, B.~AkashNarayanan, A.~Asadi, J.~M. Arrazola, U.~Azad,
  S.~Banning, C.~Blank, T.~R. Bromley, B.~A. Cordier, J.~Ceroni, A.~Delgado,
  O.~D. Matteo, A.~Dusko, T.~Garg, D.~Guala, A.~Hayes, R.~Hill, A.~Ijaz,
  T.~Isacsson, D.~Ittah, S.~Jahangiri, P.~Jain, E.~Jiang, A.~Khandelwal,
  K.~Kottmann, R.~A. Lang, C.~Lee, T.~Loke, A.~Lowe, K.~McKiernan, J.~J. Meyer,
  J.~A. Montañez-Barrera, R.~Moyard, Z.~Niu, L.~J. O'Riordan, S.~Oud,
  A.~Panigrahi, C.-Y. Park, D.~Polatajko, N.~Quesada, C.~Roberts, N.~Sá,
  I.~Schoch, B.~Shi, S.~Shu, S.~Sim, A.~Singh, I.~Strandberg, J.~Soni,
  A.~Száva, S.~Thabet, R.~A. Vargas-Hernández, T.~Vincent, N.~Vitucci,
  M.~Weber, D.~Wierichs, R.~Wiersema, M.~Willmann, V.~Wong, S.~Zhang, and
  N.~Killoran, ``Pennylane: Automatic differentiation of hybrid
  quantum-classical computations,'' {\em arXiv:1811.04968v4 [quant-ph]}, Nov
  2018.

\bibitem{refqiskit}
{Qiskit contributors}, ``Qiskit: An open-source framework for quantum
  computing,'' 2023.

\bibitem{rocha2023}
J.~C.~S. Rocha, R.~F.~I. Gomes, W.~A.~T. Nogueira, and R.~A. Dias, ``Estimating
  the number of states via the rodeo algorithm for quantum computation,'' {\em
  arXiv:2312.04322v2 [quant-ph]}, 2023.

\bibitem{ZenodoDataPublication}
{R. A. Dias, J.S.C. Rocha,W. A. Nogueira, Raphael}, ``Unraveling the rodeo
  algorithm through the zeeman model - doi: 10.5281/zenodo.10946315,'' 2024.

\bibitem{barenco1995}
A.~Barenco, C.~H. Bennett, R.~Cleve, D.~P. DiVincenzo, N.~Margolus, P.~Shor,
  T.~Sleator, J.~A. Smolin, and H.~Weinfurter, ``Elementary gates for quantum
  computation,'' {\em Phys. Rev. A}, vol.~52, pp.~3457--3467, Nov 1995.

\bibitem{bernstein1997}
E.~Bernstein and U.~Vazirani, ``Quantum complexity theory,'' {\em SIAM Journal
  on Computing}, vol.~26, no.~5, pp.~1411--1473, 1997.

\bibitem{bouland2019}
A.~Bouland, B.~Fefferman, C.~Nirkhe, and U.~Vazirani, ``On the complexity and
  verification of quantum random circuit sampling,'' {\em Nature Physics},
  vol.~15, pp.~159--163, Oct 2019.

\bibitem{beer2020book}
K.~Beer, D.~Bondarenko, T.~Farrelly, T.~Osborne, R.~Salzmann, D.~Scheiermann,
  and R.~Wolf, ``Training deep quantum neural networks,'' {\em Nature
  Communications}, vol.~11, p.~808, 02 2020.

\bibitem{schuld2021book}
M.~Schuld and F.~Petruccione, {\em Machine Learning with Quantum Computers}.
\newblock Quantum Science and Technology, Springer International Publishing,
  2021.

\bibitem{glisic2022book}
S.~Glisic and B.~Lorenzo, {\em Artificial Intelligence and Quantum Computing
  for Advanced Wireless Networks}.
\newblock Wiley, 2022.

\bibitem{combarro2023book}
E.~Combarro and S.~Gonz{\'a}lez-Castillo, {\em A Practical Guide to Quantum
  Machine Learning and Quantum Optimization: Hands-on Approach to Modern
  Quantum Algorithms}.
\newblock Packt Publishing, 2023.

\bibitem{bell1964}
J.~S. Bell, ``On the einstein podolsky rosen paradox,'' {\em Physics Physique
  Fizika}, vol.~1, pp.~195--200, Nov 1964.

\bibitem{Neumann1955book}
J.~von Neumann, {\em Mathematical Foundations of Quantum Mechanics}.
\newblock Goldstine Printed Materials, Princeton University Press, 1955.

\bibitem{fortes2015}
R.~Fortes and G.~Rigolin, ``Improving the efficiency of single and multiple
  teleportation protocols based on the direct use of partially entangled
  states,'' {\em Annals of Physics}, vol.~336, pp.~517--539, 2013.

\bibitem{mapibmqlima}
J.~John, ``Classification using quantum kernels on ibm hardware — tutorial.''
  https://medium.com/mlearning-ai/classification-using-quantum-kernels-on-ibm\\-hardware-tutorial-e2071f34fa11,
  2021.

\end{thebibliography}

\end{document}